\documentclass{article}

\usepackage{amsmath,amsthm,amsopn,amsfonts,graphicx,amssymb,dsfont,color,url}
\usepackage{gensymb} % degree symbol
\usepackage{textcomp}
\usepackage{mathrsfs}
\usepackage{color}
\usepackage{geometry}
\usepackage{subfigure}

\newcommand{\be}{\begin{equation}}
\newcommand{\ee}{\end{equation}}
\newcommand{\baa}{\begin{array}}
\newcommand{\eaa}{\end{array}}
\newcommand{\baco}{\left\{ \begin{array}}
\newcommand{\eaco}{\end{array} \right.}

\newcommand{\ds}{\displaystyle}

\def\L{\mathcal L}
\def\M{\mathcal M}
\def\Rt{\mathcal R}

\def\hD{\hat{\mu}}
\def\tA{\tilde{\alpha}}
\def\tS{\tilde{S}}
\def\tI{\tilde{I}}
\def\tR{\tilde{R}}
\def\tD{\tilde{D}}

\def\bx{\rho}

\def\Db{\overline{D}}

\definecolor{aquamarine}{rgb}{0.13, 0.68, 0.8}

\title{A parsimonious model for spatial transmission and heterogeneity in the COVID-19 propagation}

\author{Lionel Roques$^{\hbox{\small{ a,*}}}$, Olivier Bonnefon$^{\hbox{\small{ a}}}$, Virgile Baudrot$^{\hbox{\small{ a}}}$,\\ Samuel Soubeyrand$^{\hbox{\small{ a}}}$ and Henri Berestycki$^{\hbox{\small{ b}}}$ \\  \\\footnotesize{$^{\hbox{a }}$INRAE, BioSP, 84914, Avignon, France} \\\footnotesize{$^{\hbox{b }}$EHESS, CNRS, CAMS, France}
\\\footnotesize{$^{\hbox{c }}$Senior Visiting fellow, HKUST Jockey Club Institute for Advanced Study, Hong Kong University of Science and Technology}\\
\footnotesize{$^{\hbox{* }}$Contact~: lionel.roques@inrae.fr} }

\begin{document}

\maketitle

\begin{abstract}
Raw data on the cumulative number of deaths at a country level generally indicate a spatially variable distribution of the incidence of COVID-19 disease.
An important issue is to determine whether this spatial pattern is a consequence of environmental heterogeneities, such as the climatic conditions, during the course of the outbreak.
Another fundamental issue is to understand the spatial spreading of COVID-19. To address these questions, we consider four candidate epidemiological models with varying complexity in terms of initial conditions, contact rates and non-local transmissions, and we fit them to French mortality data with a mixed probabilistic-ODE approach. Using standard statistical criteria, we select the model with non-local transmission corresponding to a diffusion
on the graph of counties that depends on the geographic proximity, with time-dependent contact rate and spatially constant parameters. This original spatially parsimonious model suggests that in a geographically
middle size
 centralized country such as France, once the epidemic is established, the effect of global processes such as restriction policies, sanitary measures and social distancing overwhelms the effect of local factors. Additionally, this modeling approach reveals the latent epidemiological dynamics including the local level of immunity, and allows us to evaluate the role of non-local interactions on the future spread of the disease.
In view of its theoretical and numerical simplicity and its ability to accurately track the COVID-19 epidemic curves, the framework we develop here, in particular the non-local model and the associated estimation procedure, is  of general interest in studying spatial dynamics of epidemics.

\end{abstract}


\section{Introduction}
In France, the first cases of COVID-19 epidemic have been reported on 24 January 2020, although it appeared latter that some cases were already present in December 2019 \cite{DesBer20}. Since then, the first important clusters were observed in February in the Grand Est region and the Paris region. A few months later, at the beginning of June, the spatial pattern of the disease spread seems to have kept track of these first introductions. As this spatial pattern may also be correlated with covariates such as climate \cite{DemFle20} (see also SI~1), a fundamental question is to assess whether this pattern is the consequence of a heterogeneous distribution of some covariates or if it can be explained by the heterogeneity of the initial introduction points. In the latter case, we want to know if
the epidemic dynamics simply reflects this initial heterogeneity and can be modeled without taking into account any spatial heterogeneity in the local conditions.

SIR epidemiological models and their extensions have been proposed to study the spread of the COVID-19 epidemic at the country or state scale (e.g., \cite{RoqKle20} in France and \cite{BerFra20} in three US states), and at the regional scale  (\cite{MaiBro20,PreLiu20} in China, \cite{SalTra20} in France and \cite{GatBer20} in Italy). In all cases, a different set of parameters has been estimated for each considered region/province. One of the main goals of our study is to check whether the local mortality data at a thin spatial scale can still be well explained by a single set of parameters, at the country scale, but spatially varying initial conditions. In this study we consider SIR models with time-dependent contact rate, to track changes over time in the dynamic reproductive number as in the branching process considered in \cite{BerFra20}. Using standard statistical criteria, we compare four types of models, whose parameters are either global at the country scale or spatially heterogeneous, and with non-local transmission or not. We work with mortality data only, as these data appear to be more reliable and less dependent on local testing strategies than confirmed cases, in settings where cause of death is accurately determined \cite{GarCha20}. Additionally, it was already shown that for this type of data, SIR models outperform other classes of models (an SEIR models and a branching process) in the three US states considered in \cite{BerFra20}.  The approach we develop here is in line with the general principle of using parsimonious models eloquently emphasized in \cite{BerFra20}.

Another important issue that such models may help to address is the quantification of the relative effects of restrictions on inter-regional travels, {\it vs} reductions in the probability of infection per contact at the local scale. France went into lockdown on March 17, 2020, which was found to be very effective in reducing the spread of the disease. It divided the effective reproduction number (number $\Rt_t$ of secondary cases generated by an infectious individual) by a factor 5 to 7 at the country scale by May 11 \cite{RoqKle20b,SalTra20}. This is to be compared with the estimate of the basic reproduction number $\Rt_0$ carried out in France at the early onset of the epidemic, before the country went into lockdown (with values $\Rt_0=2.96$ in \cite{SalTra20} and $\Rt_0=3.2$ in \cite{RoqKle20}). Contact-surveys data \cite{ZhaLit20} for Wuhan and Shanghai, China, have found comparable estimates of the reduction factor. The national lockdown in France induced important restrictions on movement, with e.g. a mandatory home confinement except for essential journeys, leading to a reduction of the number of contacts. In parallel,  generalized mask wearing and use of hydroalcoholic gels reduced the probability of infection per contact.

After the lockdown, restrictions policies were generally based on raw data rather than modeling. An efficient regulation at the local scale would need to know the current number of infectious and the level of immunity at the scale of counties. The French territory is divided into administrative units called `d\'epartements', analogous to counties. We use 'counties' in the sequel for d\'epartements. These quantities cannot be observed directly in the absence of large scale testing campaigns  or spatial random sampling. In particular, there is a large number of unreported cases. Previous studies developed a mixed mechanistic-probabilistic framework to estimate these quantities at the country scale. These involved  estimating the relative probability of getting tested for an infected individual {\it vs} a healthy individual, leading to a factor x8 between the number of confirmed cases and the actual number of cases before the lockdown \cite{RoqKle20}, and x20 at the end of the lockdown period \cite{RoqKle20b}. This type of framework -- often referred to as
`mechanistic-statistical modeling' -- aims at connecting the solution of continuous state models such as differential equations with complex data, such as noisy discrete data, and identifying latent processes such as the epidemiological process under consideration here. Initially introduced for  physical models and data \cite{Ber03}, it is becoming standard in ecology \cite{SouLai09}.

Using this framework the objectives of our study are (i) to assess whether the spatial pattern observed in France is due do some local covariates or is simply the consequence of the heterogeneity in the initial conditions together with global processes at the country scale; (ii)
to evaluate the role of non-local interactions on the spread of the disease; (iii) to propose a tool for real-time monitoring the main components of the disease in France, with a particular focus on the local level of immunity.

\section*{Materials and Methods}

\subsection*{Data} Mainland France (excluding Corsica island) is made of 94 counties called 'd\'epartements'. The daily number of hospital deaths -- excluding nursing homes -- at the county scale are available from Santé Publique France since 18 March 2020 (and available as Supplementary Material). The daily number of observed deaths (still excluding nursing homes) in county $k$ during day $t$ is denoted by $\hD_{k,t}$.

We denote by  $[t_i,t_f]$ the observation period and by $n_d$ the number of considered counties. To avoid a too large number of counties with 0 deaths at initial time, the observation period ranges from $t_i=$ March~30 to $t_f=$ June~11, corresponding to $n_t=74$ days of observation. All the counties of mainland France (excluding Corsica) are taken into account, but the Ile-de-France region, which is made of 8  counties with a small area is considered as a single geographic entity. This leads to $n_d=87$.

\subsection*{Mechanistic-statistical framework\label{sec:framework}}

The mechanistic-statistical framework is a combination of a mechanistic model that describes the epidemiological process, a probabilistic observation model and a statistical inference procedure. We begin with the description of four mechanistic models, whose main characteristics are given in Table~\ref{table:defmodels}.

\subsubsection*{Mechanistic models}

\paragraph{Model~$\M_{0}$: SIR model for the whole country.} The first model is the standard mean field SIRD model that was used in~\cite{RoqKle20,RoqKle20b}:
\begin{equation}
\baco{l} \label{eq:modela}
 \ds S'(t)=- \frac{\alpha(t)}{N} \, S \, I, \vspace{1mm}\\
 \ds I'(t)= \frac{\alpha(t)}{N} \, S \, I - (\beta+\gamma) \, I, \vspace{1mm}\\
 \ds R'(t)=\beta \,  I,\vspace{1mm}\\
 \ds D'(t)=\gamma \, I,
\eaco
\end{equation}
with $S$ the susceptible population, $I$ the infectious population, $R$ the recovered population, $D$ the number of deaths due to the epidemic and $N$ the total population, in the whole country.  For simplicity, we assume that $N$ is constant, equal to the current population in France, thereby neglecting the effect of the small variations of the population on the coefficient $\alpha(t)/N$. The parameter $\alpha(t)$ is the contact rate (to be estimated)  and $1/\beta$ is the mean time until an infectious becomes recovered. The results in \cite{ZhoYu20} show that the median period of viral shedding is 20 days, but the infectiousness tends to decay before the end of this period: the results in \cite{HeLau20} indicate that infectiousness starts from 2.5 days before symptom onset and declines within 7 days of illness onset. Based on these observations we assume here that $1/\beta=10$ days. The parameter $\gamma$ corresponds to the death rate of the infectious. It was estimated independently in \cite{RoqKle20,RoqKle20b} and \cite{SalTra20}, leading to a value $\gamma=5 \cdot 10^{-4}$. This value only takes into account the deaths at hospital, and is therefore consistent with the data that we used here.

\paragraph{Model~$\M_{1}$: SIR model at the county scale with globally constant contact rate and no spatial transmission.} The model $\M_{0}$ is applied at the scale of each county $k$, leading to compartments $S_k$, $I_k$, $R_k$, $D_k$ that satisfy an equation of the form \eqref{eq:modela}, with $N$ replaced by $N_k$, the total population in the county $k$. In this approach, the contact rate $\alpha(t)$ is assumed to be the same in all of the counties.

% \begin{equation}
% \baco{l} \label{eq:model0} \tag{$\M_{1}$}
%  \ds S_k'(t)=- \frac{\alpha(t)}{N_k} \, S_k \, I_k, \vspace{1mm}\\
%  \ds I_k'(t)= \frac{\alpha(t)}{N_k} \, S_k \, I_k - (\beta+\gamma) \, I_k, \vspace{1mm}\\
%  \ds R_k'(t)=\beta \,  I_k,\vspace{1mm}\\
%  \ds D_k'(t)=\gamma \, I_k,
% \eaco
% \end{equation}

\begin{table}
\begin{center}
    \begin{tabular}{|c|c|c|c|c|c|}
  \hline
  & Heterog.   & Heterog. & Intercounty  &  Nb. parameters \\
    & initial data   & contact rate &  transmission &  \\
    \hline
  $\M_{0}$ & no  & no & no & $ n_t$ \\
  \hline
  $\M_{1}$ & yes & no & no & $ n_t$  \\
   \hline
  $\M_{2}$ & yes  & yes   & no &  $n_d \times n_t$   \\
  \hline
  $\M_{3}$ & yes  & no   & yes & $n_t+2$ \\
  \hline
  \end{tabular}
  \caption{Main characteristics of the four models. The quantity $n_t=74$ corresponds to the number of days of the observation period and $n_d=87$ corresponds to the number of administrative units.}
  \label{table:defmodels}
\end{center}
\end{table}

\paragraph{Model $\M_{2}$:  SIR model at the county scale with spatially heterogeneous contact rate and no spatial transmission.} With this approach, the model $\M_{1}$ is extended by assuming that the contact rate $\alpha_k(t)$ depends on the considered county.

\paragraph{Model $\M_{3}$: County scale model with globally constant contact rate and spatial transmission.} The model $\M_{1}$ is extended to take into account disease transmission events between the counties:
\begin{equation}
\baco{l} \label{eq:model3}
 \ds S_k'(t)=- \frac{\bx(t)}{N_k} \, S_k \, \sum_{j=1}^{n_d} w_{j,k}\, I_j, \vspace{1mm}\\
 \ds I_k'(t)=\frac{\bx(t)}{N_k} \, S_k \, \sum_{j=1}^{n_d} w_{j,k}\, I_j - (\beta+\gamma) \, I_k , \vspace{1mm}\\
 \ds R_k'(t)=\beta \,  I_k,\vspace{1mm}\\
 \ds D_k'(t)=\gamma \, I_k.
\eaco
\end{equation}
\noindent The weights $w_{j,k}$ describe the dependence of the contagion  rates with respect to the distance between counties. We assume a power law decay with the distance:
\begin{equation}
    w_{j,k}=\frac{1}{1+(\hbox{dist}(j,k)/d_0)^{\delta}},
\end{equation}
with $\hbox{dist}(j,k)$ the geographic distance (in km) between the centroids of counties $j$ and $k$, $d_0>0$ a proximity scale, and $\delta>0$. Thus, the model involves two new global parameters, $d_0$ and $\delta$, compared to model $\M_1$. This model extends the  Kermack-McKendrick SIR model to take into account non-local spatial interactions. It was introduced by Kendall \cite{Ken57} in continuous variables. The model we adopt here is inspired from the study of \cite{BonBer18} in a different context  where the same types of weights have been used. We thus take into account diffusion on the weighted graph of counties in France. This amounts to considering that individuals in a given county are infected by individuals form other counties with a probability that decreases with distance as a power law, in addition to contagious individuals from their own county.  This dependence of social spatial interactions with respect to the distance is supported, notably, by  \cite{BroHuf06} that analyzed the short-time dispersal of bank notes in the US. We also refer to \cite{MeyHel14} for a thorough discussion on the various applications of power law dispersal kernels since they were introduced by Pareto \cite{Par1896}.

With this non-local contagion model, in contradistinction to epidemiological models with dispersion such as reaction-diffusion epidemiological models \cite{GauGha10}, the movements of the individuals are not modeled explicitly. The model implicitly assumes that infectious individuals may transmit the disease to susceptible individuals in other counties, but eventually return to their county of origin.
This has the advantage of avoiding unrealistic changes in the global population density.

\subsubsection*{Observation model}
We denote by $\Db_k(t)$ the expected cumulative number of deaths given by the model, in county $k$. With the mean-field model $\M_{0}$, we assume that it is proportional to the population size: $\Db_k(t)=D(t) \, N_k/N,$ with $N_k$ the population in county $k$ and $N$ the total French population.
With models~$\M_{1},$ $\M_{2}$ and $\M_{3}$, we simply have $\Db_k(t)=D_k(t)$. The expected daily increment in the number of deaths given by the models in a county $k$ is $\Db_k(t)-\Db_k(t-1)$.

The observation model assumes that the daily number of new observed deaths $\hD_{k,t}$ in county $k$ follows a Poisson distribution with mean value $\Db_k(t)-\Db_k(t-1)$:
\begin{equation}\label{eq:model_poisson}
    \hD_{k,t}\sim \text{Poisson}(\Db_k(t)-\Db_k(t-1)).
\end{equation}
Note that the time $t$ in the mechanistic models is a continuous variable, while the observations  $\hD_{k,t}$ are reported at discrete times. For the sake of simplicity, we used the same notation $t$ for the days in both the discrete and continuous cases. In the formula \eqref{eq:model_poisson} $\Db_k(t)$ (resp. $\Db_k(t-1)$) is computed at the end of day $t$ (resp. $t-1$).

\subsection*{Initial conditions}

In models $\M_{1}$, $\M_{2}$, $\M_{3}$, at initial time $t_i$, we assume that the number of susceptible cases is equal to the number of inhabitants in county $k$:  $S_k(t_i)=N_k$, the number of recovered is $R(t_i)=0$ and the number of deaths is given by the data: $D_k(t_i)=\hD_{k,t_i}$. To
 initialise the number of infectious, we use the equation $D'(t)=\gamma \, I(t)$, and we define $I(t_i)$ as $1/\gamma$ $\times$ (mean number of deaths over the period ranging from $t_i$ to $29$ days after $t_i$):
\begin{equation}\label{eq:IO}
 I_k(t_i)=\frac{1}{\gamma}\frac{1}{30} \sum\limits_{s=t_i,\ldots,t_i+29}\hD_{k,s}.
\end{equation}
The 30-days window was chosen such that there was at least one infectious case in each county. In model~$\M_{0}$, the initial conditions are obtained by adding the initial conditions of model~$\M_{1}$ (or equivalently, $\M_{2}$) over all the counties.

\subsection*{Statistical inference}

\paragraph{Real-time monitoring of the parameters and data assimilation procedure.} To smooth out the effect of small variations in the data, and to avoid identifiability issues due to the large number of parameters, while keeping the temporal dependence of the parameters, the parameters $\alpha(t)$ and $\alpha_k(t)$  of the ODE models $\M_{0}$, $\M_{1}$, $\M_{2}$ are estimated by fitting auxiliary problems with time-constant parameters over moving windows $(t-\tau/2,t+\tau/2)$ of fixed duration equal to $\tau$ days. These auxiliary problems are denoted respectively by $\tilde{\M}_{0,t}$, $\tilde{\M}_{1,t}$, and $\tilde{\M}_{2,t}$ (see SI~2 for a precise formulation of these problems). The initial conditions associated with this system, at the date $t-\tau/2$ are computed iteratively from the solution of $\M_{0}$, $\M_{1}$ and $\M_{2}$, respectively.

\paragraph{Inference procedure.} For simplicity, in all cases, we denote by
$$f_{\Db_k,\hD_k}(s):= \frac{(\Db_k(s)-\Db_k (s-1))^{\hD_{k,s }}}{\hD_{k,s}!}\,  e^{-(\Db_k(s)-\Db_k(s-1))}$$the probability mass function associated with the observation process~\eqref{eq:model_poisson} at date $s$ in county $k$, given the expected cumulative number of deaths $\Db_k$ given by the considered model in county $k$.

In models $\tilde{\M}_{0,t}$ and $\tilde{\M}_{1,t}$, the estimated parameter is $\tA$. The likelihood $\mathcal{L}$ is defined as the probability of the observations (here, the increments $\{\hD_{k,s}\}$) conditionally on the parameter. Using the assumption that the increments $\hD_{k,s}$ are independent conditionally on the underlying SIRD process~ $\tilde{\M}_{0,t}$ (resp. $\tilde{\M}_{1,t}$), we get:
\begin{align*}
 \mathcal{L}(\tA)\ds & :=P(\{\hD_{k,s}, \, k=1,\ldots,n_d, \, s=t-\tau/2,\dots,t+\tau/2\} |\tA) \\ &
      =\prod_{k=1}^{n_d}\prod_{s=t-\tau/2}^{t+\tau/2} f_{\Db_k,\hD_k}(s).
\end{align*} We denote by $\tA_{t}^*$ the corresponding maximum likelihood estimator, and we set $\alpha(t)=\tA_{t}^* \hbox{ in model }\M_{0} \hbox{ (resp. }\M_{1}).$

For model $\tilde{\M}_{2,t}$,
the inference of the parameters $\tA_k$ is carried out independently in each county, leading to the likelihoods:
\begin{align*}
\mathcal{L}_k(\tA_k)\ds &:=P(\{\hD_{k,s}, \ s=t-\tau/2,\dots,t+\tau/2\} |\tA_k)\\
      &=\prod_{s=t-\tau/2}^{t+\tau/2} f_{\Db_k,\hD_k}(s).
\end{align*}
We denote by $\tA_{k,t}^*$ the corresponding maximum likelihood estimator, and we set $\alpha_k(t)=\tA_{k,t}^*$  in model $\M_{2}.$

For model $\M_{3}$, we apply a two-stage estimation approach. We first use the estimate obtained with model $\M_1$ by setting
    $\bx(t)= C \, \alpha(t),$
where $\alpha(t)$ is the estimated contact rate of model $\M_1$ and $C$ is a constant (to be estimated; note that estimating $\rho(t)$ means estimating $n_t$ parameters). Thus, given $\alpha(t)$, the only parameters to be estimated are the constant $C$, the proximity scale $d_0$ and the exponent $\delta$. They are estimated by maximizing:
\begin{align*}
    \label{eq:likelihoodEDP}
    \mathcal{L}(C,d_0,\delta)\ds & :=P(\{\hD_{k,s}, \ k=1,\ldots,n_d, \, s=t_i,t_f\}  |C,d_0,\delta) \\
    & =\prod_{k=1}^{n_d}\prod_{s=t_i}^{t_f}f_{\Db_k,\hD_k}(s).
\end{align*}

\

\paragraph{Model selection.}  We use the Akaike information criterion (AIC) \cite{Aka74} and the Bayesian information criterion (BIC) \cite{Sch78} to compare the models. For both criteria, we need to compute the likelihood function associated with the model, with the parameters determined with the above inference procedure:
\begin{equation}
    \L(\M_m)=\prod_{k=1}^{n_d}\prod_{s=t_i}^{t_f}f_{\Db_k,\hD_k}(s),
\end{equation}
with $m=0\,,1,\,2,\,3$ and $\Db_k$ the expected (cumulative) number of deaths in county $k$ given by model $(\M_m)$. Given the number of parameters $\sharp \M_m$ estimated in model $\M_m$, the AIC score is defined as follows:
$$AIC(\M_m)=2\, \sharp\M_m -2 \, \ln(\L(\M_m)),$$
and the BIC score:
$$BIC(\M_m)=\sharp\M_m\, \ln(K)-2 \, \ln(\L(\M_m)),$$with $K=n_d \times n_t=6\,438$ the number of data points.

\

\noindent \textit{Numerical methods}.
To find the maximum likelihood estimator, we used a BFGS constrained minimization algorithm, applied to $ - \ln (\L)$  via the Matlab \textsuperscript{\tiny \textregistered} function \textit{fmincon}. The ODEs were solved thanks to a standard numerical algorithm, using Matlab\textsuperscript{\tiny\textregistered} \textit{ode45} solver. The Matlab \textsuperscript{\tiny \textregistered} codes are available as Supplementary Material.

\section*{Results}

\begin{figure*}
\center
%\graphicspath{{images/}}
\includegraphics[width=0.32\textwidth]{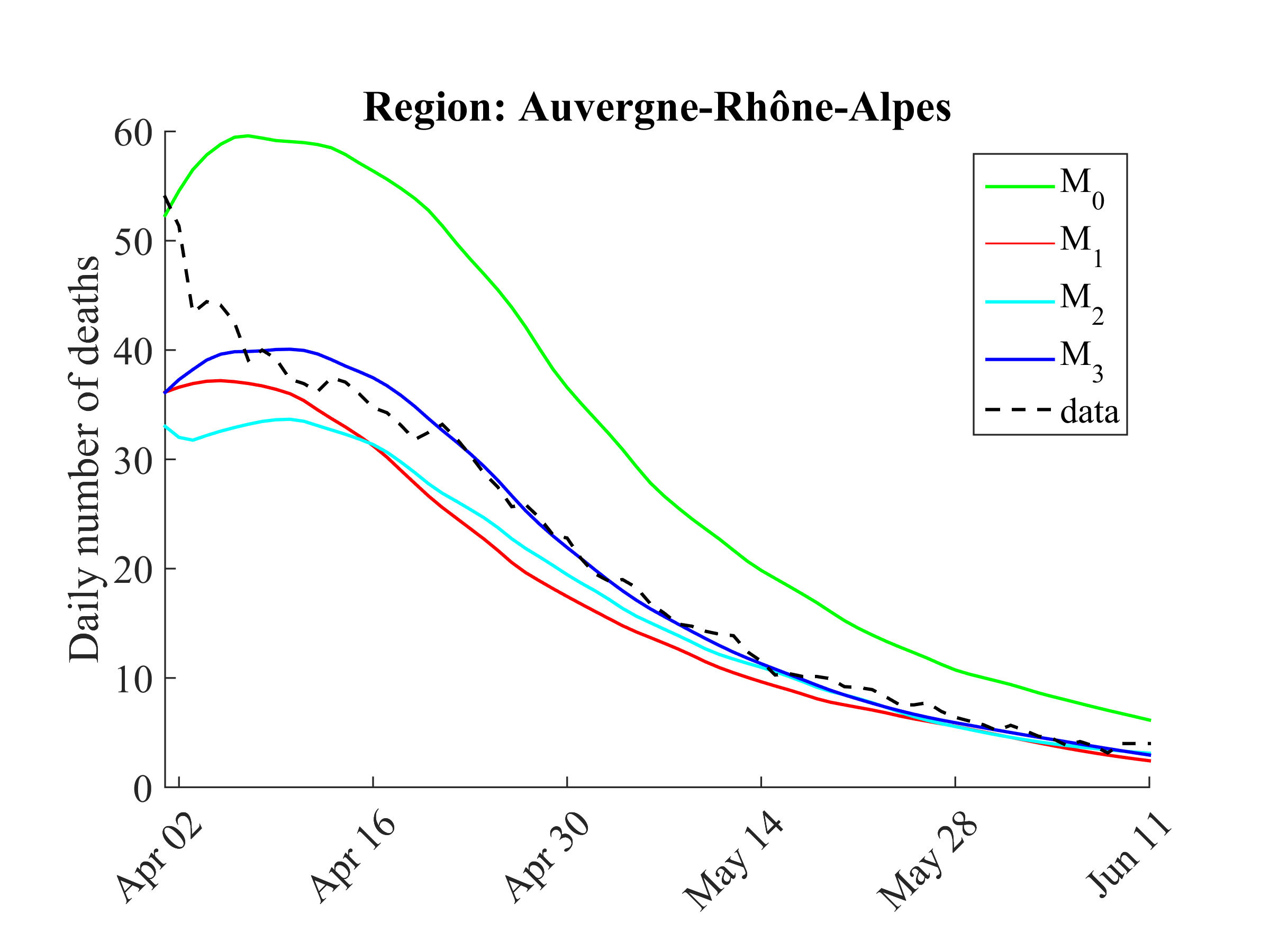}
\includegraphics[width=0.32\textwidth]{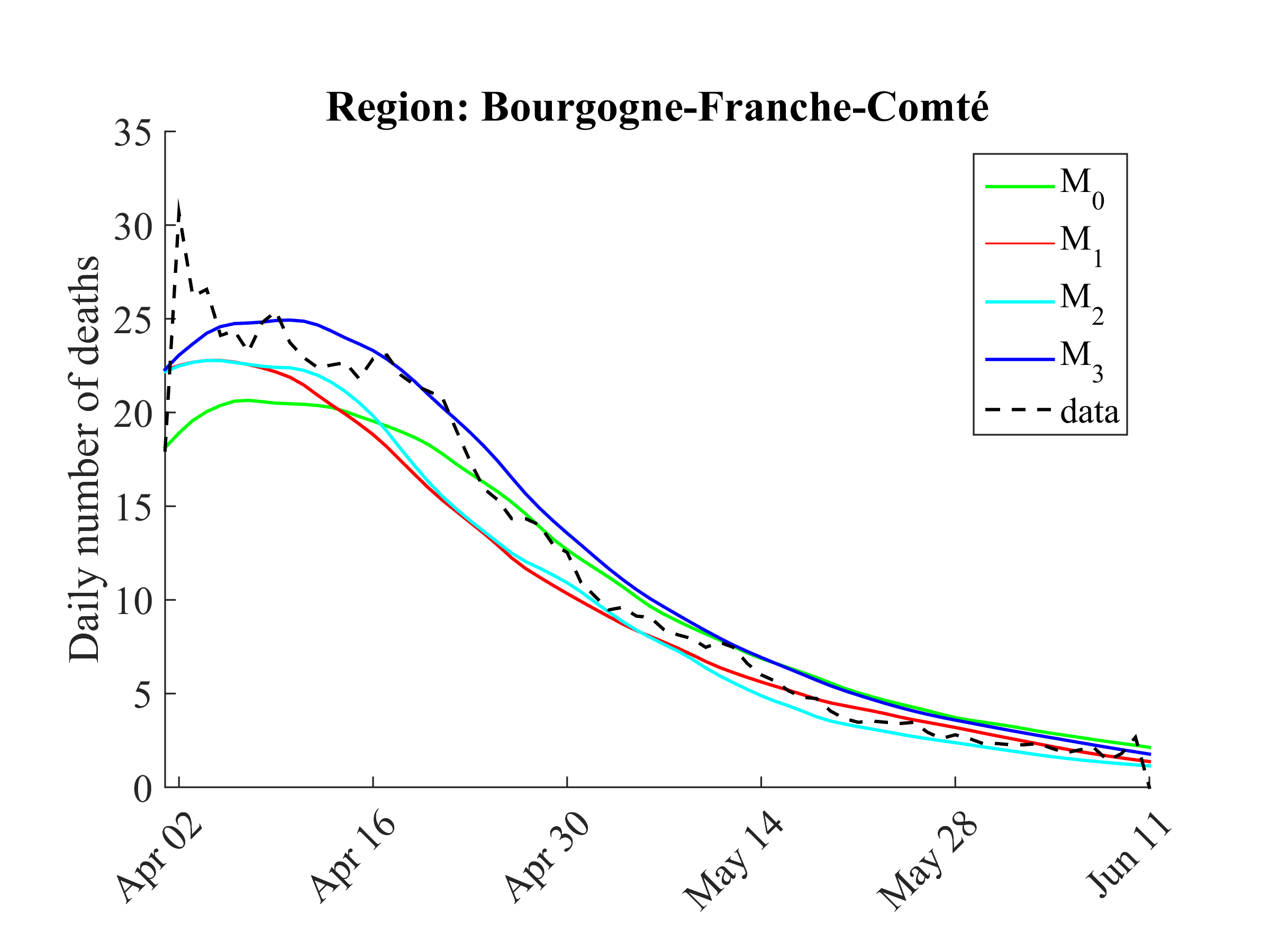}
\includegraphics[width=0.32\textwidth]{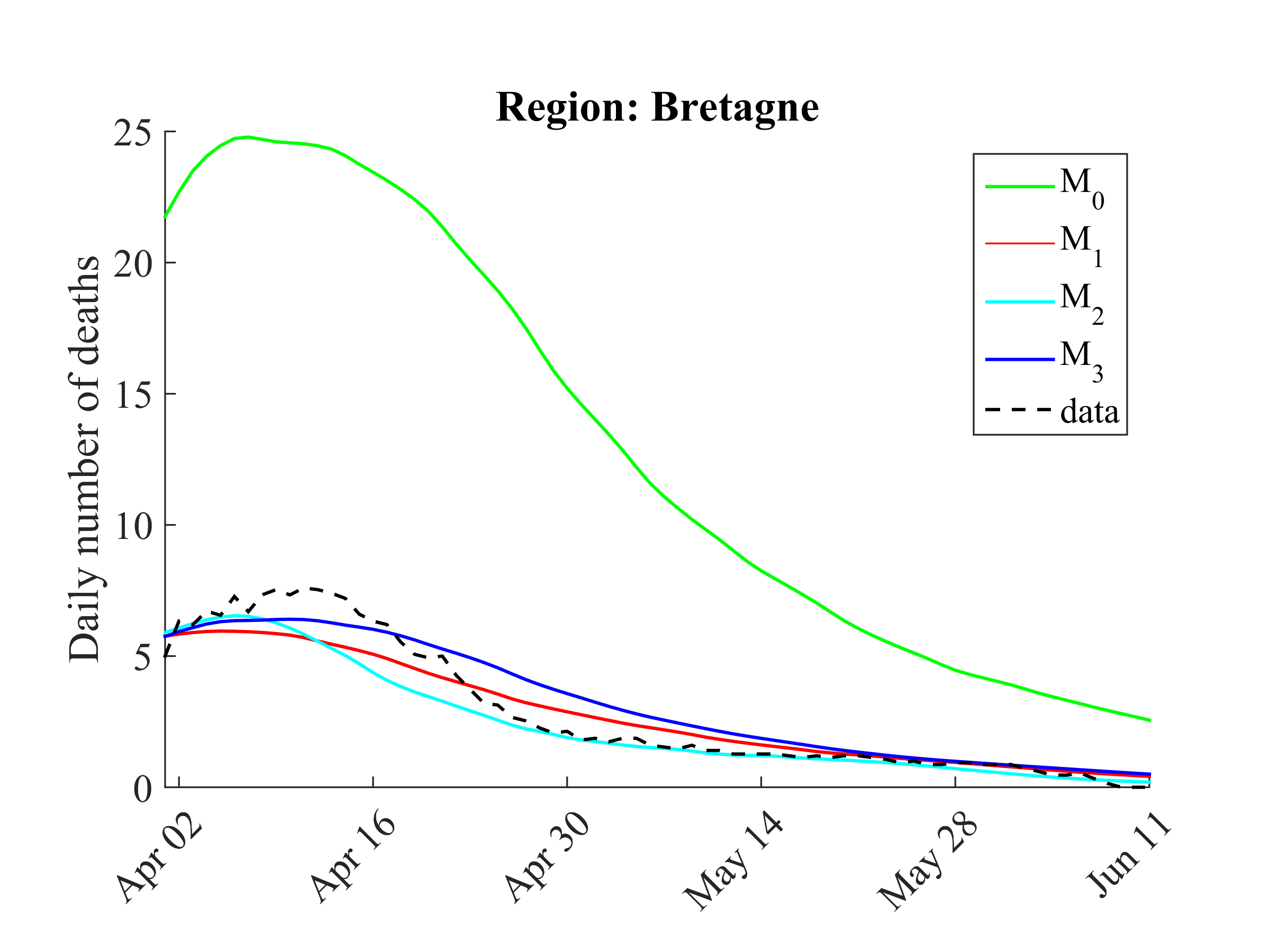}
\includegraphics[width=0.32\textwidth]{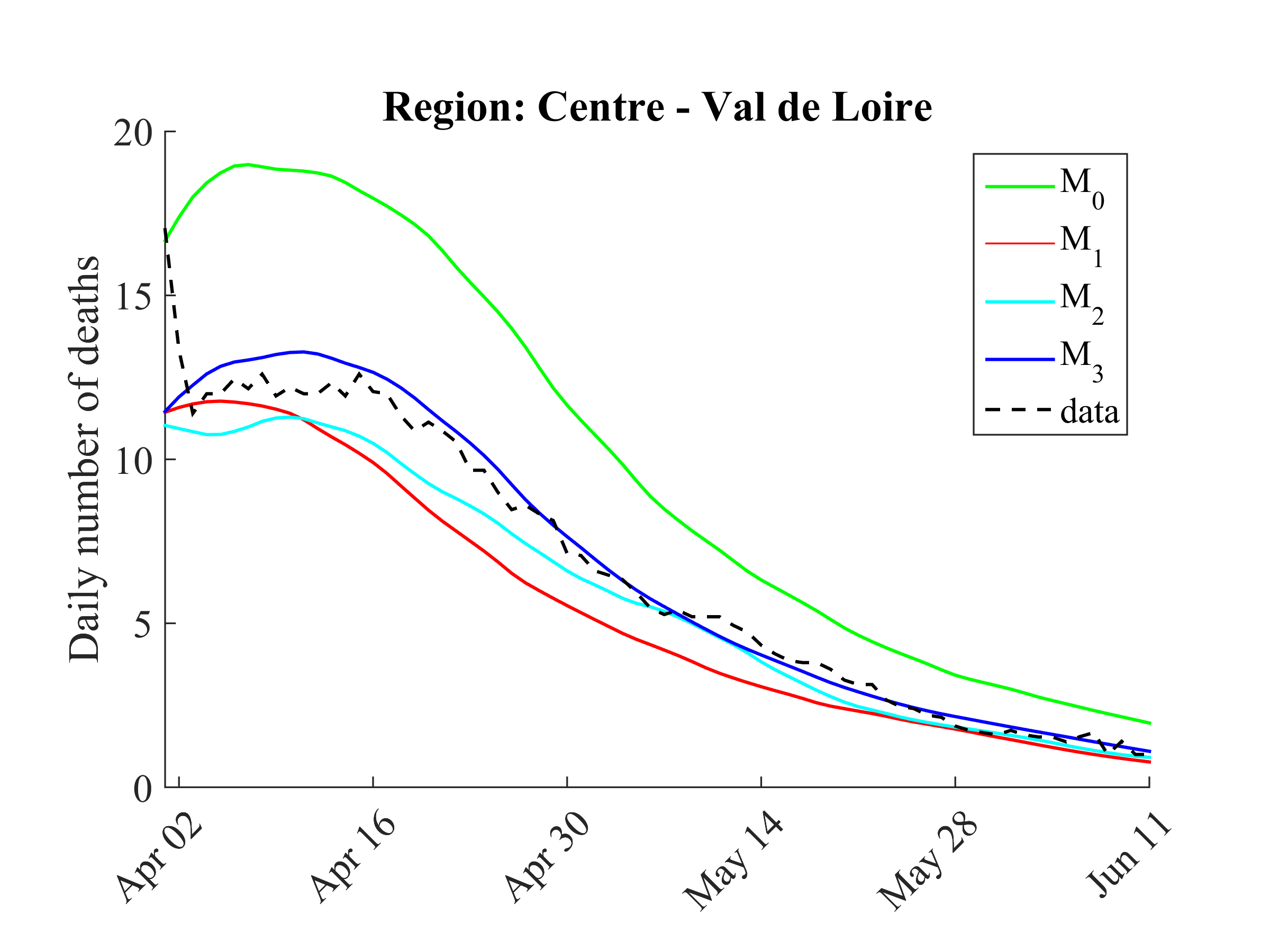}
\includegraphics[width=0.32\textwidth]{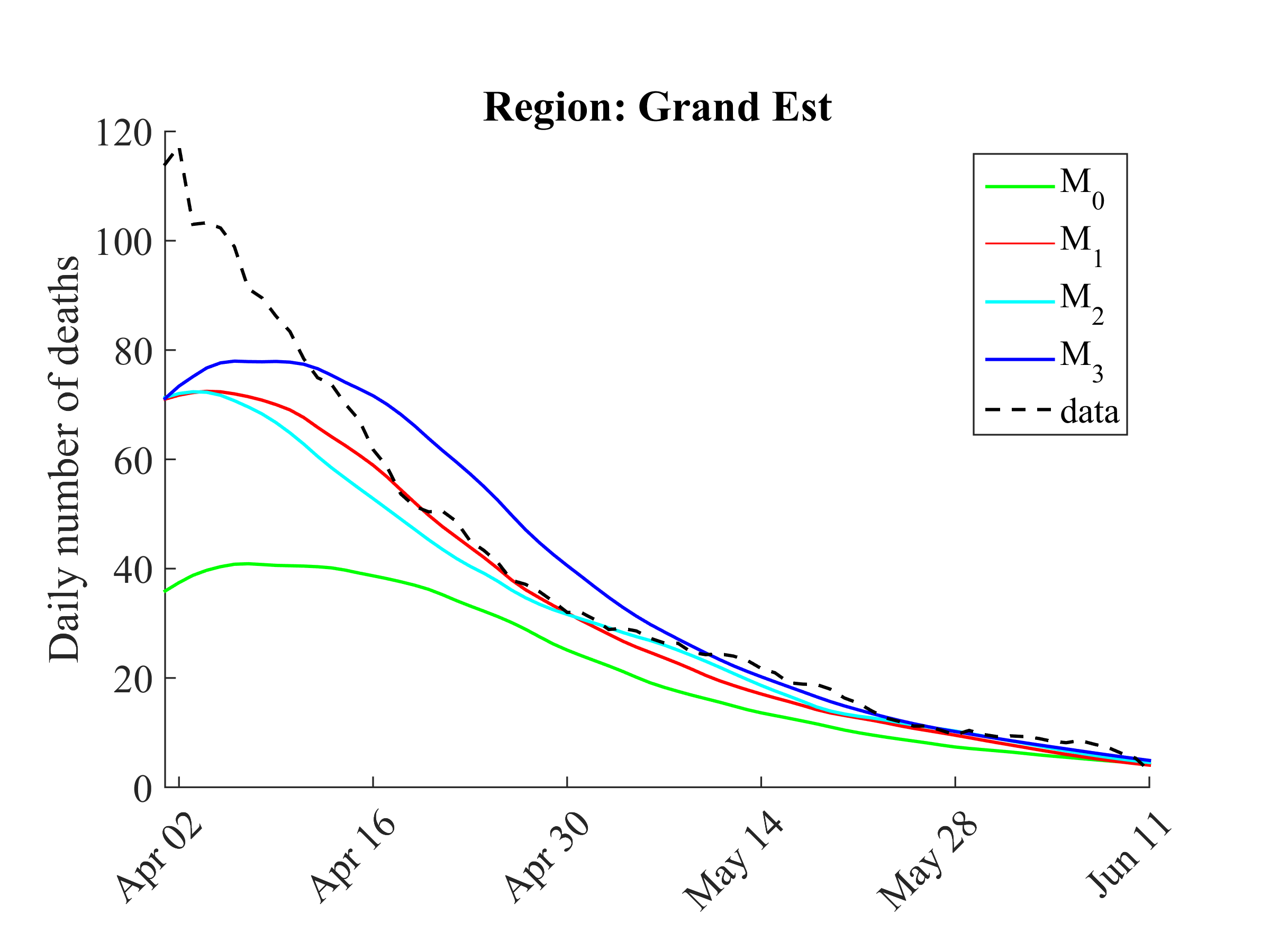}
\includegraphics[width=0.32\textwidth]{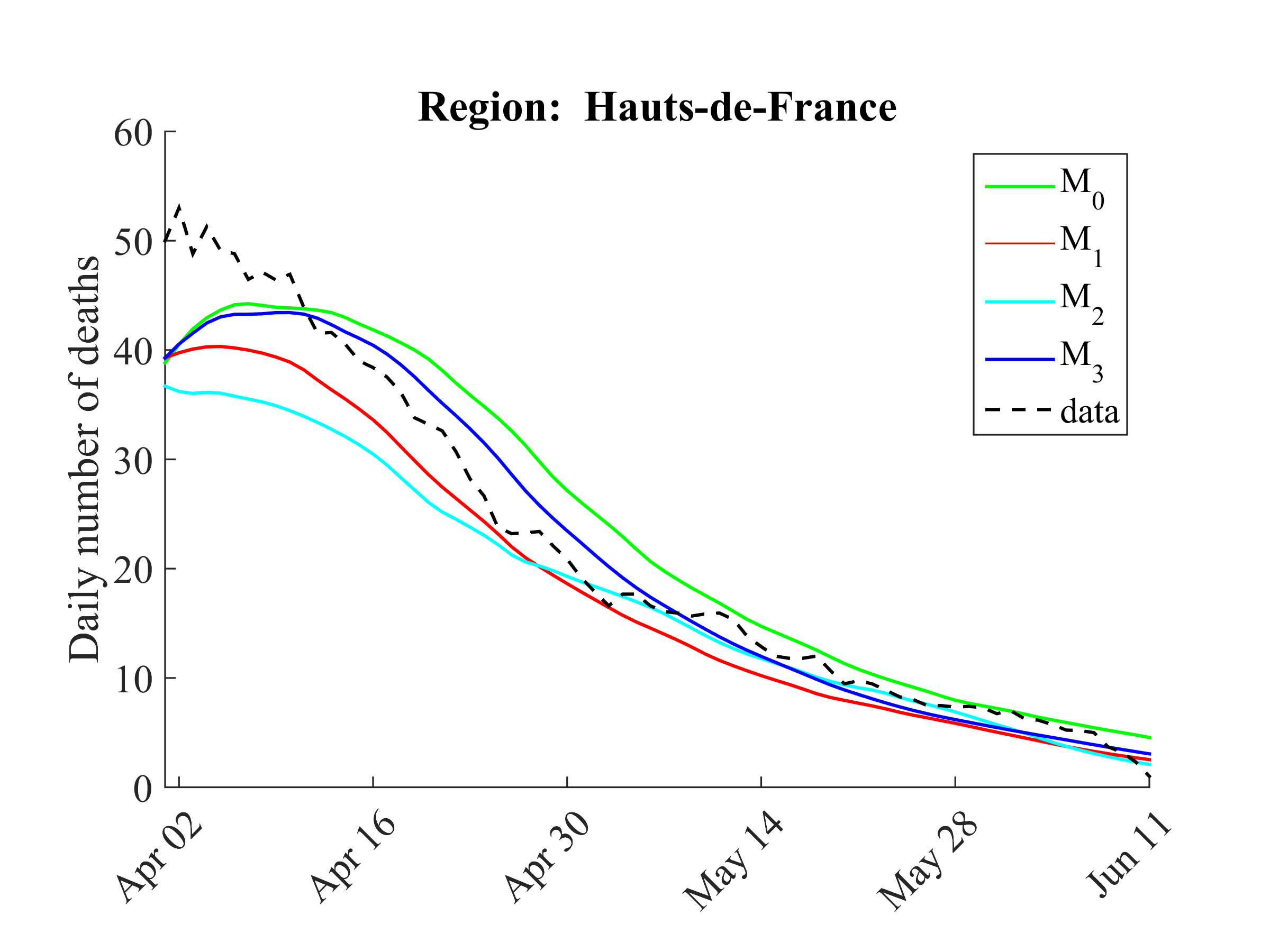}
\includegraphics[width=0.32\textwidth]{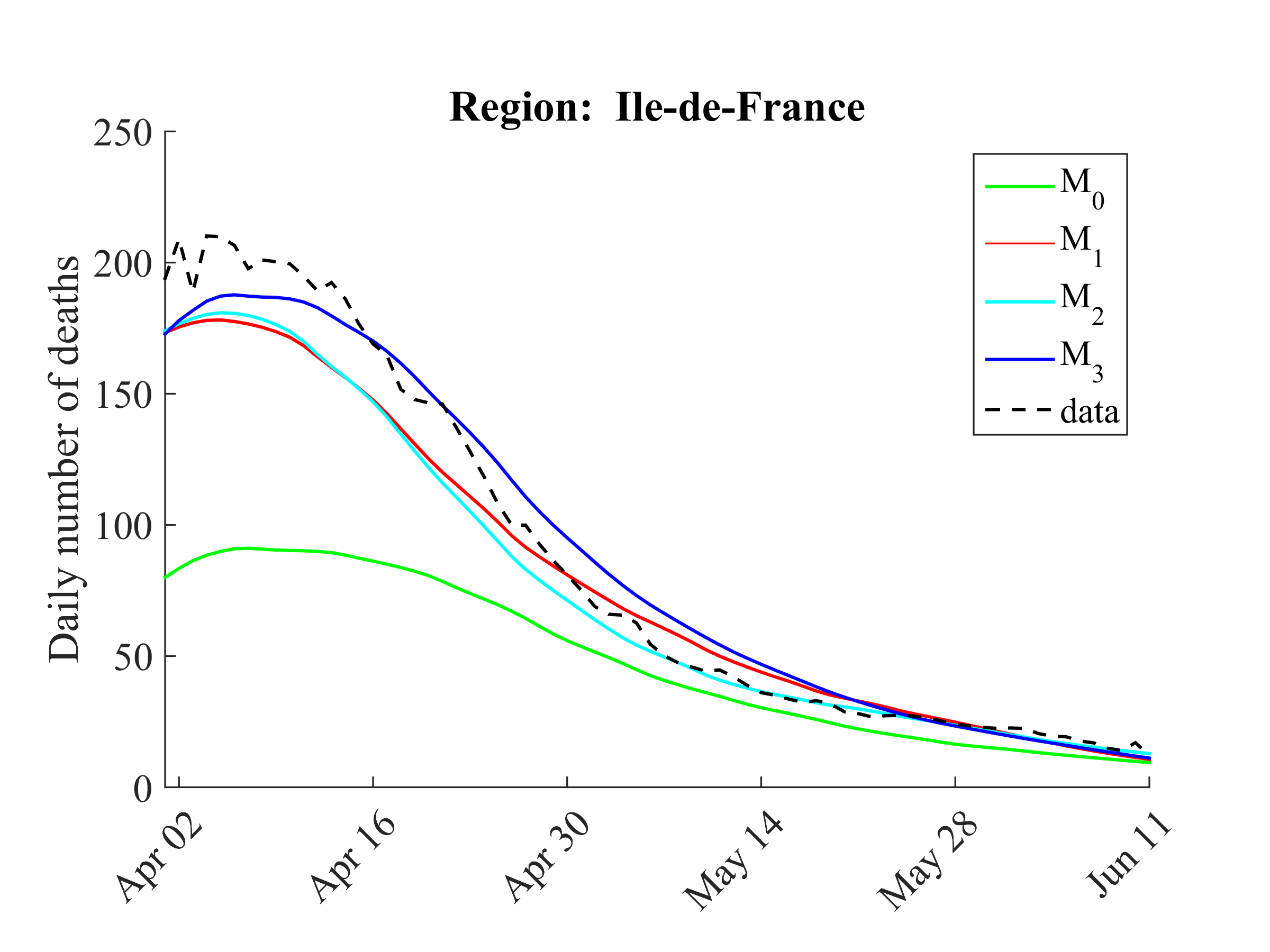}
\includegraphics[width=0.32\textwidth]{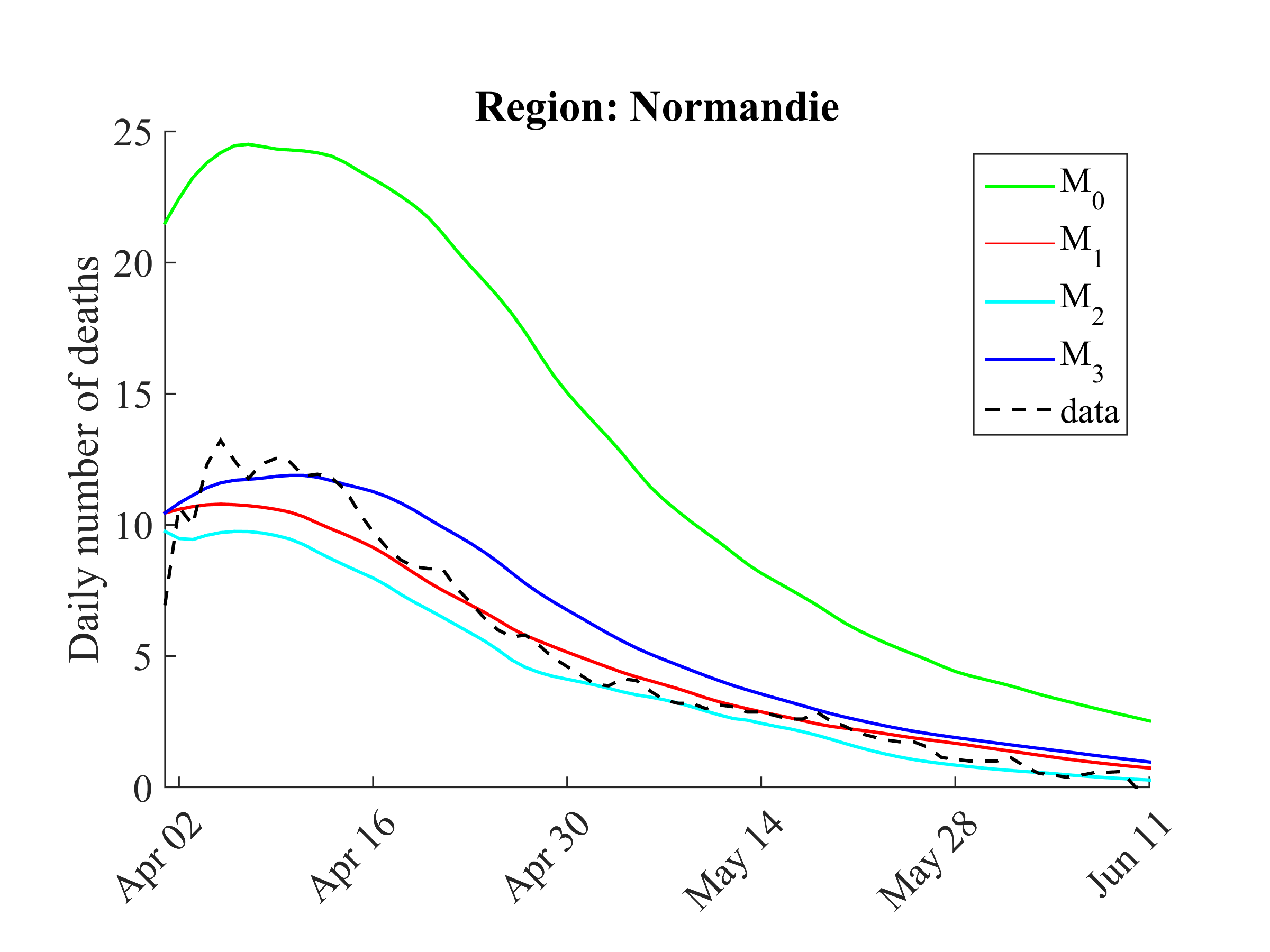}
\includegraphics[width=0.32\textwidth]{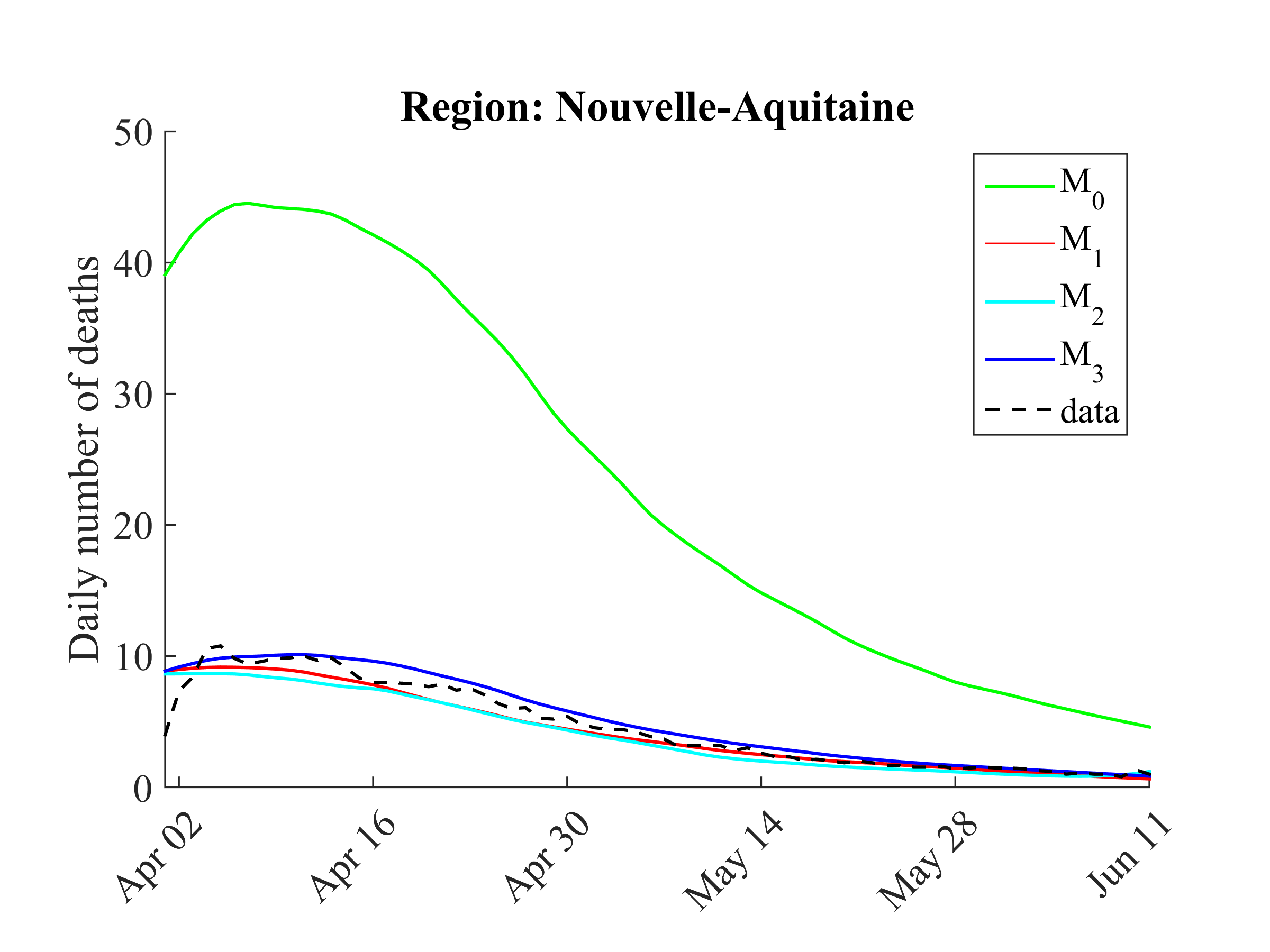}
\includegraphics[width=0.32\textwidth]{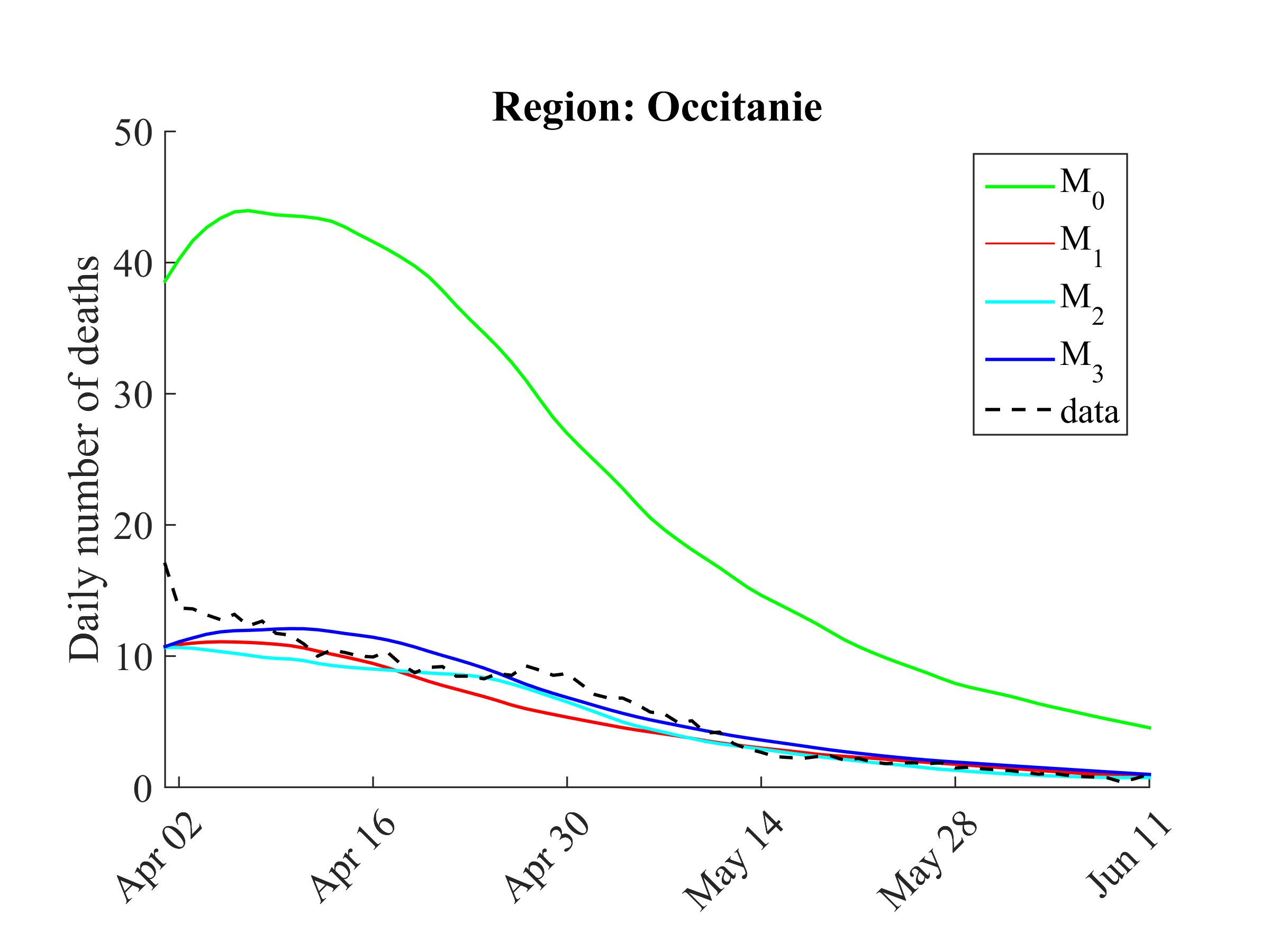}
\includegraphics[width=0.32\textwidth]{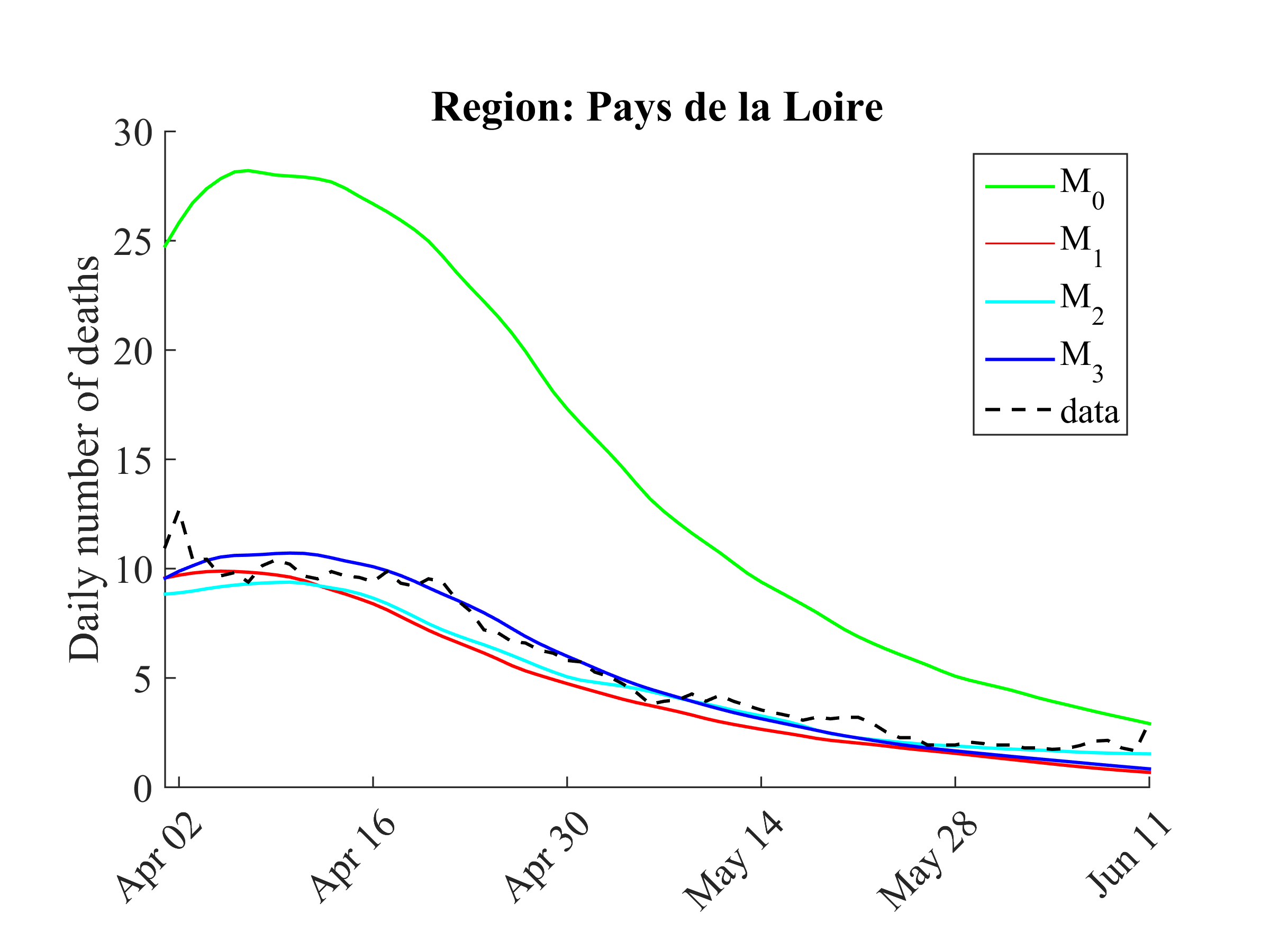}
\includegraphics[width=0.32\textwidth]{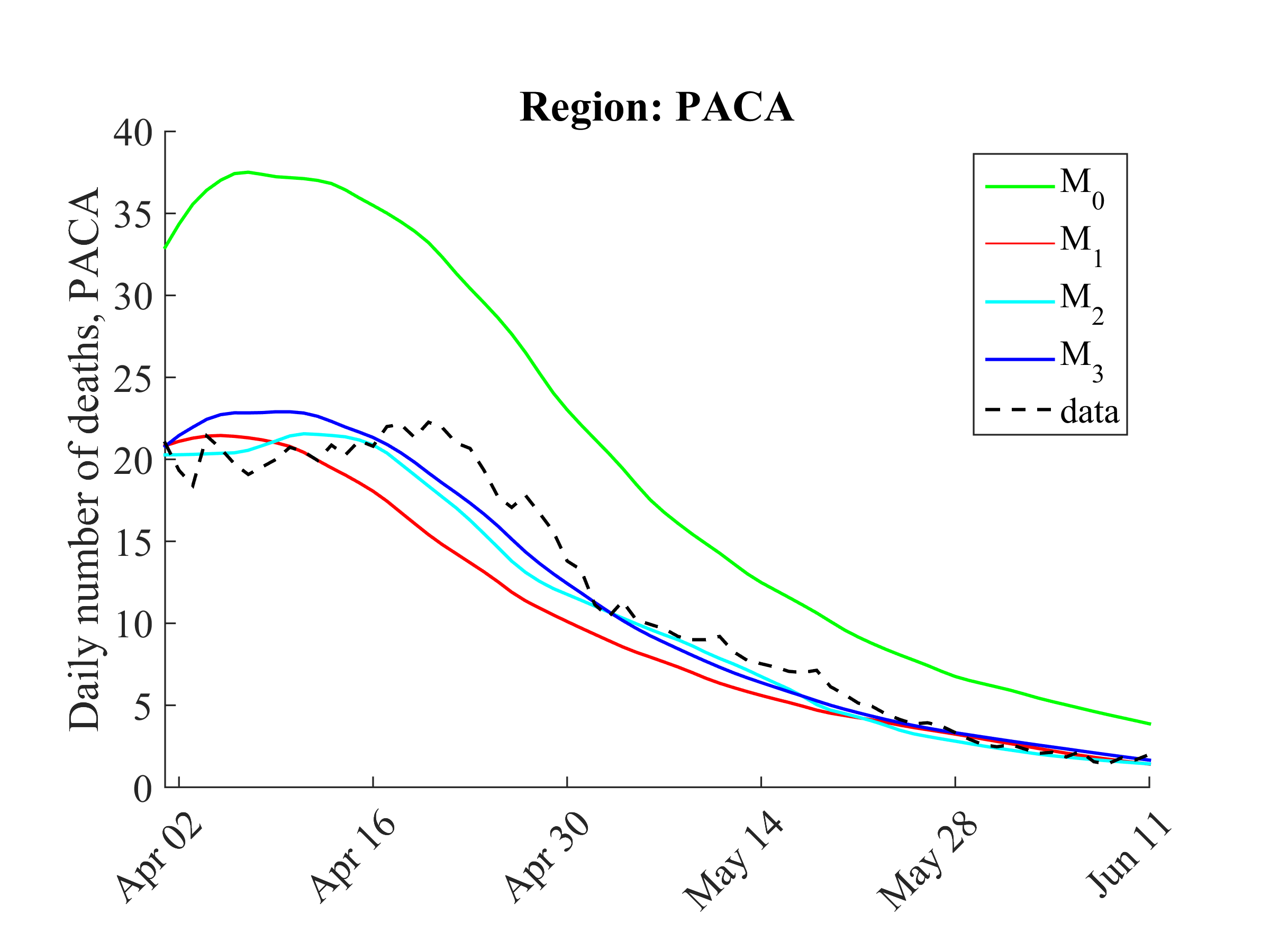}
\caption{Model fit at the regional scale. The data have been smoothed (moving average over 15 days), for easier graphical comparison with the models.}
\label{fig:model_fit_region}
\end{figure*}

\noindent \textit{Model fit and model selection.} To assess model fit, we compared the daily increments in the number of deaths given by each of the four models with the data. For each model, we used the parameters corresponding to the maximum likelihood estimators. The comparisons are carried out
at the regional scale: mainland France is divided into 12 administrative regions, each of which is made of several counties. The results are presented in Fig.~\ref{fig:model_fit_region}. In all the regions, the models $\M_1,$ $\M_2$ and $\M_3$ lead to a satisfactory visual fit of the data, whereas the mean field model $\M_0$ does not manage to reproduce the variability of the dynamics among the regions.

The log-likelihood, AIC and BIC values are given in Table~\ref{table:AIC}. Models $\M_1$, $\M_2$, $\M_3$ lead to significantly higher likelihood values than model $\M_0$. This reflects the better fit obtained with these three models, compared to model $\M_0$ and shows the importance of taking into account the spatial heterogeneities in the initial densities of infectious cases. On the other hand, the log-likelihood, though higher with model $\M_2$ is close to that obtained with $\M_1$, and the model selection criteria are both strongly in favor of model $\M_1$. This shows that the spatial heterogeneity in the contact rate does not have a significant effect on the epidemic dynamics within mainland France.

Model $\M_3$ with spatial transmission leads to an intermediate likelihood value, between those of models $\M_1$ and $\M_2$, with only $2$ additional parameters with respect to model $\M_1$. As a consequence, the model selection criteria exhibit a strong evidence in favor of the selection of model $\M_3$. This means a large part of the difference between models $\M_1$ and $\M_2$ can be captured by taking into account the spatial transmission, which therefore seems to have a significant effect on the epidemic dynamics.

\begin{figure}
\center
%\graphicspath{{images/}}
\includegraphics[width=0.5\textwidth]{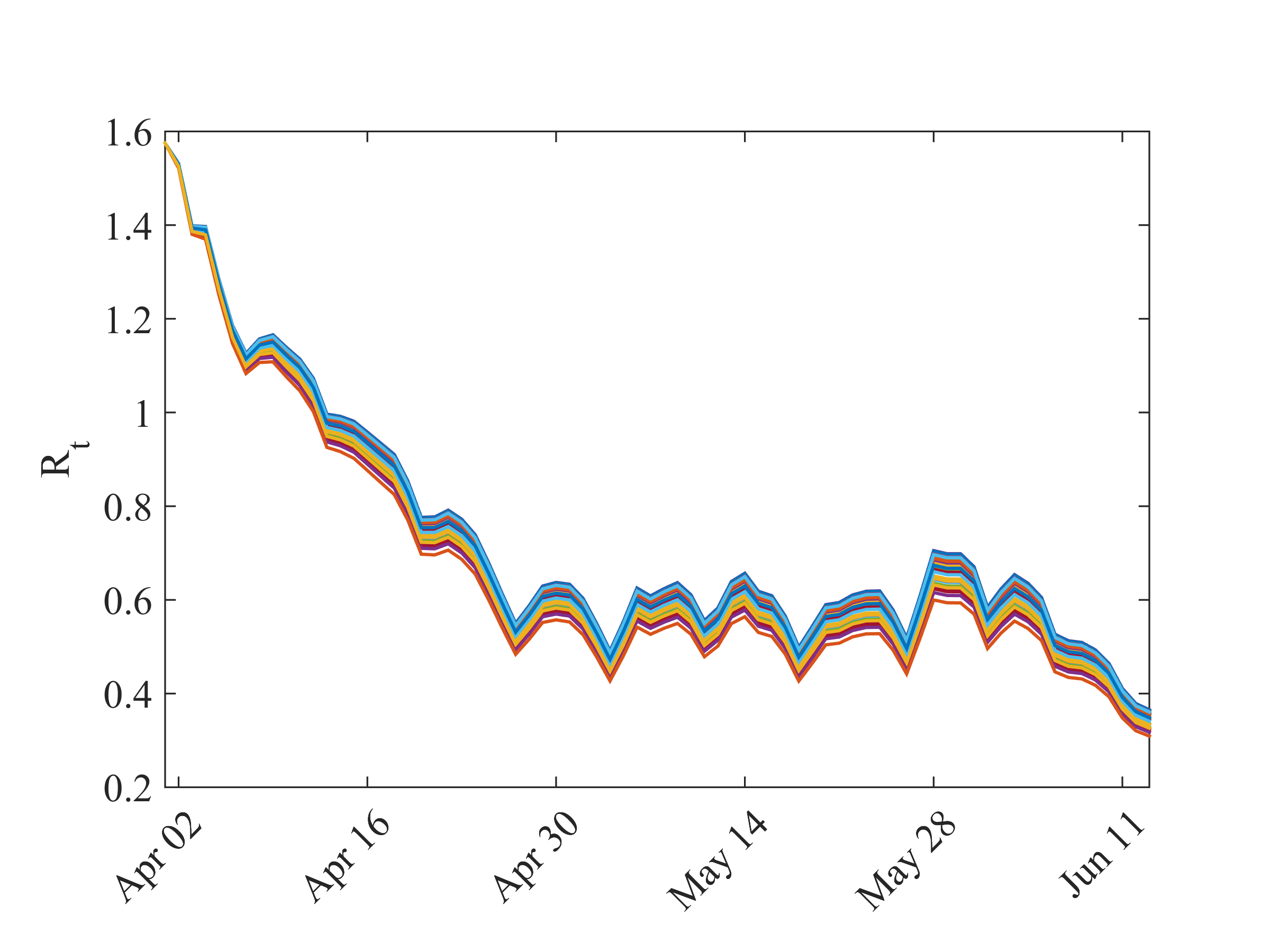}
\caption{Dynamics of the effective reproduction rate $\Rt_{t}^k$ given by model $\M_1$ over the 87 considered counties.}
\label{fig:Rt}
\end{figure}

As a byproduct of the estimation of the parameter $\alpha(t)$ (resp. $\alpha_k(t)$) of model $\M_1$ (resp. $\M_2$), we get an estimate of the effective reproduction number in each county, which is given by the formula \cite{NisGer09}:$$\Rt_{t}^k=\frac{\alpha(t)}{\beta+\gamma}\frac{S_k(t)}{N_k},$$so that $I_k'(t)<0$ (the epidemic tends to vanish) whenever $\Rt_{t}^k<1$, whereas $I_k'(t)>0$ whenever $\Rt_{t}^k>1$ (the number of infectious cases in the population follows an increasing trend). The dynamics of $\Rt_{t}^k$ obtained with model $\M_1$ are depicted in Fig.~\ref{fig:Rt}, clearly showing a decline in $\Rt_{t}^k$, as already observed in \cite{RoqKle20b,SalTra20}, but there the computation was at a fixed date.

For model $\M_3$, the maximum likelihood estimation gives $C=0.87$, $d_0=2.16~$km and $\delta=1.85$, which yields a nearly quadratic decay of the weights with the distance. The value of $d_0$, indicates that non-local contagion plays a secondary role compared to within-county contagion: the minimum distance between two counties is 36~km, leading to a weight of $5.5/1000$, to be compared with the weight 1 for within-county contagion. However, the fact that the parameter $C$ is significantly smaller than $1$ (recall that the contact rate in model $\M_3$ is $\rho(t)=C\, \alpha(t)$ with $\alpha(t)$ the contact rate in model $\M_2$) shows that the non-local contagion term plays an important role on the spreading of the epidemic.

\

\begin{figure}
\center
\includegraphics[width=.8\linewidth]{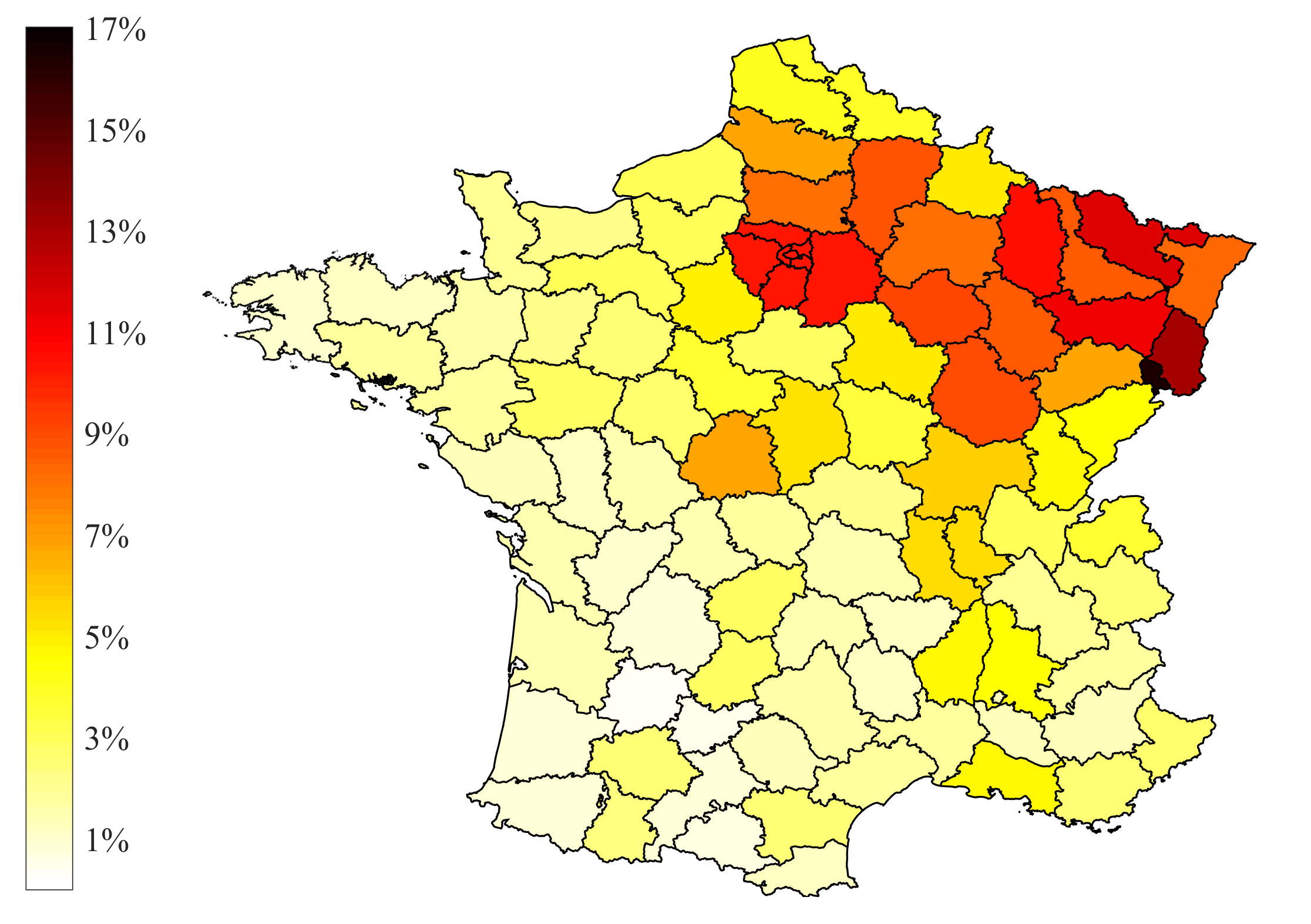}
\caption{Estimated immunity rate, at the county scale, by June 11, 2020, using model $\M_3$. }
\label{fig:immunity}
\end{figure}

\noindent \textit{Immunity rate.} Using model $\M_3$, which leads to the best fit, we derive the number of recovered individuals (considered here as immune) at a date $t$, in each county, and the immunity rate $R_k/N_k$. It is presented at time $t_f$ in Fig.~\ref{fig:immunity}. The full timeline of the dynamics of immunity obtained with model $\M_3$ since the beginning of April is available as Supplementary Material (see SI~1).

\begin{table}
\begin{center}
    \begin{tabular}{|c|c|c|c|c|}
  \hline
  Model  & AIC & BIC & Log-likelihood & $\Delta$AIC \\
    \hline
 $\M_{0}$ & $2.68 \cdot 10^4$ & $2.73 \cdot 10^4$ & $-13.4 \cdot 10^3$ & $-9.57 \cdot 10^3$\\
  \hline
  $\M_{1}$ &$ 1.74\cdot 10^4$ & $ 1.79\cdot 10^4$ & $-8.62 \cdot 10^3$ & $-220$\\
  \hline
  $\M_{2}$ & $2.97\cdot 10^4$& $ 7.36 \cdot 10^4$& $-8.41 \cdot 10^3$ & $-1.25 \cdot 10^4$ \\
  \hline
  $\M_{3}$ & $ 1.72\cdot 10^4$ & $ 1.77\cdot 10^4$ & $-8.52 \cdot 10^3$ & 0 \\
  \hline
  \end{tabular}
  \caption{Log-likelihood, AIC and BIC values for the four models. The last column $\Delta$AIC corresponds to the difference with the AIC value of the best model (here $\M_3$).}
  \label{table:AIC}
 \end{center}
\end{table}

\

%%%%%%%%%%%%%%%%%%%%%%%%%%%%%%%%%%%

\noindent \textit{Limiting movement vs limiting the probability of transmission per contact.} Before the lockdown, the basic reproduction number $\Rt_0$ in France was about 3 \cite{SalTra20,RoqKle20}, and was then reduced by a factor 5 to 7, leading to values around 0.5 (see \cite{RoqKle20b,SalTra20} and Fig.~\ref{fig:Rt}). This corresponds to a contact rate $\alpha(t)\approx \beta \, \Rt_t\approx 0.3$ before the lockdown and $\alpha(t)\approx 0.05$ after the lockdown in model $\M_1$. Let us now consider a hypothetical scenario of a new outbreak with an effective reproduction number that rises again to reach values above 1. Due to the higher awareness of the population with respect to epidemic diseases and the new sanitary behaviors induced by the first COVID-19 wave, the reproduction number will probably not reach again values as high as 3.

\begin{figure}
\center
\includegraphics[width=.8\linewidth]{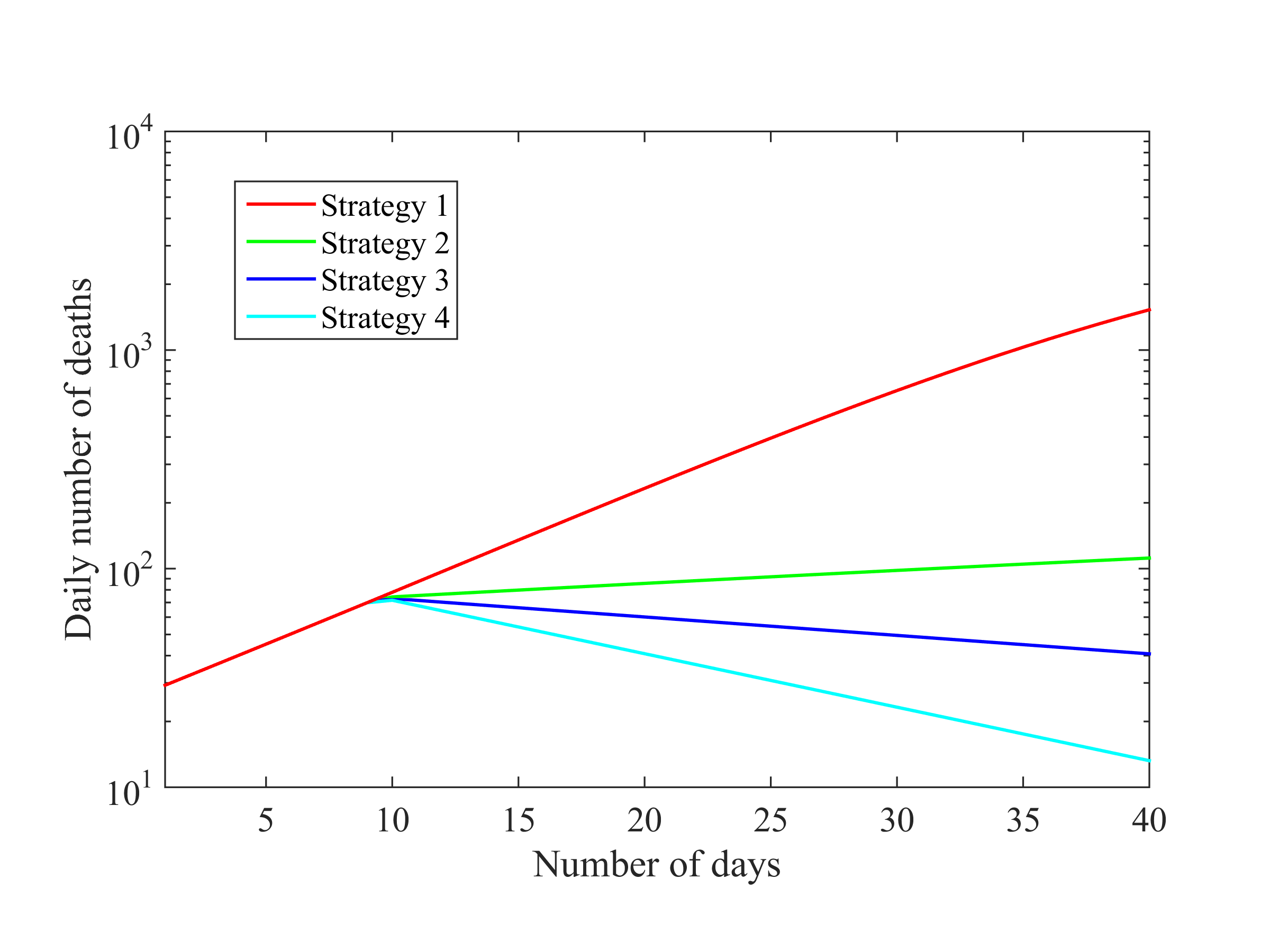}
\caption{Daily number of deaths due to a new outbreak in logarithmic scale; comparison between four management strategies. The number of deaths is computed over the whole country.}
\label{fig:scenarios}
\end{figure}

The new outbreak scenario is described as follows: we start from the state of the epidemic at June 11, and we assume a 'local' contact rate  $\rho(t)$  jumping to the value $0.11$ in model $\M_3$ (corresponding to a 2-fold increase compared to the previous 30 days). In parallel, to describe the lifting of restrictions on individual movements, we set $d_0=20$~km for the proximity scale in model $\M_3$. This new outbreak runs during 10 days, and then, we test four strategies:
\begin{itemize}
    \item[-] Strategy~1: no restriction. The parameters remain unchanged: $\rho(t)=0.11$ and $d_0=20$~km;
    % calcul du Rt: I' = (\beta +\gamma) I (R_t-1)
    % curve fitting 1.4741e+05*exp(0.1*(R-1)*(x-10))
    \item[-] Strategy~2: restriction on intercounty movement. The parameter $\rho(t)=0.11$ is unchanged, but $d_0=2.16$~km, corresponding to its estimated value during the period $(t_i,t_f)$;
    \item[-] Strategy~3: reduction of the contact rate within each county (e.g., by wearing masks), but no restriction on intercounty movement: $\rho(t)=0.05$ and $d_0=20$~km;
    \item[-] Strategy~4: reduction of the contact rate within each county and restriction on intercounty movement: $\rho(t)=0.05$ and $d_0=2.16$~km.
\end{itemize}
The daily number of deaths corresponding to each scenario is presented in Fig.~\ref{fig:scenarios}. We estimate in each case a value of the effective reproduction number $\Rt_t$ over the whole country by fitting the global number of infectious cases with an exponential function over the last 30 days. As expected, the more restrictive the strategy, the less the number of deaths. After 30 days, the cumulative number of deaths with the first strategy is $17\, 271$, and $\Rt_t\approx 2$. Restriction on intercounty movement (strategy 2) leads to a 81\% decrease in the cumulative number of deaths ($3\, 281$ deaths) and $\Rt_t\approx 1.2$; reducing the contact rate within each county leads to a 88\% decrease (strategy~3, $2\, 139$ deaths) and $\Rt_t\approx 0.8$; finally, control strategy 4, which combines both types of restrictions leads to a 91\% decrease ($1\, 503$ deaths) and $\Rt_t\approx 0.4$.

%%%%%%
\section*{Discussion}

%* Discuss the main result:
We show here that a parsimonious model
%involving country-wide parameters
can reproduce the local dynamics of the COVID-19 epidemic in France with a relatively high goodness of fit. This is achieved despite the spatial heterogeneity across French counties of some environmental factors potentially influencing the disease propagation.
%For COVID-19, despite the spatial heterogeneity across French counties of some environmental factors potentially influencing disease propagation, the local epidemic dynamics can be explained with a relatively high goodness of fit, using a parsimonious model.
Indeed, our model only involves the initial spatial distribution of the infectious cases and spatially-homogeneous (i.e. countrywide) parameters.  For instance, the mean temperature during the considered period ranges from $12.0\degree$C to $18.4\degree$C depending on the region. We do observe a negative correlation  between the mean temperature and the immunity rate (see Fig.~S1), however it does not reflect a causality. Actually, our study shows that if there is an effect of local covariates such as the mean temperature on the spread of the disease once it emerged, its effect  is of lower importance compared to the global processes at work at the country scale. Such covariates might play a major role in the emergence of the disease, but our work focuses on the disease dynamics after the emergence.

Hence, we find that initial conditions and spatial diffusion are the main drivers of the spatial pattern of the COVID-19 epidemic. This result may rely on specific circumstances: e.g., mainland France covers a relatively middle-sized area, with mixed urban and countryside populations across the territory, a relatively homogeneous population age distribution, and a high level of centralism for public decision (in particular regarding the disease-control strategy). Of course, these features are not universal.
In other countries with more socio-environmental diversity within which environmental drivers and state decentralization could significantly induce spatial variations in disease spread and, consequently, in which countrywide parameters would not be appropriate.
Moreover, at the global scale, the COVID-19 dynamics in different countries seem highly contrasted as illustrated by the data of Johns Hopkins University Center for Systems Science and Engineering \cite{DonDu20} ---see also \url{http://covid19-forecast.biosp.org/} \cite{SouRib20}--- and could probably not be explained with a unique time-varying contact rate parameter.
Nonetheless, this model could be adapted to many other situations at an appropriate geographical level, with a single well defined political decision process.

% *
Herd immunity requires that a fraction $1-1/\Rt_0\approx 70\%$ of the population has been infected. It is far from being reached at the country scale in France, but we observe that the fraction of immune individuals strongly varies across the territory, with possible local immunity effects. For instance, in the most impacted county the immunity rate is 16\%, whereas it is less than $1$\% in less affected counties. At a thinner grain scale, even higher rates may be observed, for instance,  by April~4 the proportion of people with confirmed SARS-CoV-2 infection based on antibody detection was  41\% in a high-school located in Northern France \cite{FonTon20}.

Real-time monitoring of the immunity level will be crucial to define efficient management policies, if a new outbreak occurs. We propose such a tool which is based on the modeling approach $\M_3$ of this paper (see Immunity tab in  \url{http://covid19-forecast.biosp.org/}).  Remarkably, the estimated levels of immunity are comparable to those observed in Spain by population-based serosurveys \cite{PolPer20}, with values ranging from $0.5\%$ to $13.0\%$ at the beginning of May at the provincial resolution. Such a large-scale serological testing campaign has not yet been carried out in France. However, if such data become available in France, our predictions could be evaluated and our model updated accordingly by including this new dataset in our estimation procedure. The mechanistic-statistical approach that we followed here can indeed be easily adapted to multiple observation processes.

%* discuss the results on the different mitigation strategies.
Our results indicate that ---at the country scale--- travel restriction alone, although they may have a significant effect on the cumulative number of deaths and the reproduction number over a definite period, are less efficient than social distancing and other sanitary measures. Obviously, these results may strongly depend on the parameter values, which have been chosen here on the basis of values estimated during the lockdown period. This is consistent with results for China \cite{ChiDav20}, where travel restrictions to and from Wuhan have been shown to have a modest effect unless paired with other public health interventions and behavioral changes.

The model selection criteria led to a strong evidence in favor of the selection of model $\M_3$ with non-local transmission and spatially-constant contact rate. It is much more parsimonious than the fully heterogeneous model $\M_2$ and is therefore better suited to isolating key features of the epidemiological dynamics \cite{BerFra20}. Despite important restrictions on movement during the considered period (mandatory home confinement except for essential journeys until 11 May and a 100~km travel restriction until 2 June), the model $\M_3$ was also selected against the model $\M_1$ which does not take into account non-local transmission. This shows that intercounty transmission is one of these key-features that the non-local model manages to take into account.

More generally, in France just as in Italy \cite{GatBer20}, the spatial pattern of COVID-19 incidence indicates that spatial processes play a key role. At this stage, only a few models can address this aspect. Some have adopted a detailed spatio-temporal modeling approach and use mobility data (see \cite{GatBer20} for an SEIR-like model with 9 compartments). The framework we develop here, including the non-local model and the associated estimation procedure, should be of broad interest in studying the spatial dynamics of epidemics, due to its theoretical and numerical simplicity and its ability to accurately track the epidemics. This approach applies when geographic distance matters, which may not be the case at the scale of countries like the US. However, it does at more regional scales. Furthermore, we can envision natural extensions of the approach that would take into account long range dispersal events in the interaction term.
%by modifying the distance function.

%%%%%%
\section*{Data availability}
All data used in this manuscript are publicly
available. French mortality data at the county scale are available at https://www.gouvernement.fr/info-coronavirus/carte-et-donnees
and are also available as Supplementary Material.

\section*{Acknowledgments}
This work was funded by INRAE (MEDIA network) and EHESS. We thank Jean-Fran\c cois Rey for assistance in developing the web app.

% Bibliography
%\bibliography{biblio_lionel_jab_drive}

%\bibliography{biblio_lionel_jab_drive}

\begin{thebibliography}{10}

\bibitem{DesBer20}
A Deslandes, et~al., {SARS-COV-2 was already spreading in France in late
  December 2019}.
\newblock {\em\protect\textit{International Journal of Antimicrobial
  Agents}}, 106006 (2020).

\bibitem{DemFle20}
J Demongeot, Y Flet-Berliac, H Seligmann, Temperature decreases spread
  parameters of the new {Covid}-19 case dynamics.
\newblock {\em\protect\textit{Biology}} \textbf{9}, 94 (2020).



\bibitem{RoqKle20}	
L Roques, E Klein, J Papaix, A Sar, S Soubeyrand, {Using early data to estimate
  the actual infection fatality ratio from COVID-19 in France}.
\newblock {\em\protect\textit{Biology}} \textbf{9}, 97 (2020).

\bibitem{BerFra20}
AL Bertozzi, E Franco, G Mohler, MB Short, D Sledge, The challenges of modeling
  and forecasting the spread of COVID-19.
\newblock {\em\protect\textit{Proceedings of the National Academy of
  Sciences}} https://doi.org/10.1073/pnas.2006520117 (2 July 2020).

\bibitem{MaiBro20}
BF Maier, D Brockmann, Effective containment explains subexponential growth in
  recent confirmed COVID-19 cases in China.
\newblock {\em\protect\textit{Science}} \textbf{368}, Issue 6492, 742--746 (2020).

\bibitem{PreLiu20}
K Prem, et~al., {The effect of control strategies to reduce social mixing on
  outcomes of the COVID-19 epidemic in Wuhan, China: a modelling study}.
\newblock {\em\protect\textit{The Lancet Public Health}} \textbf{5}, Issue 5, 261--270 (2020).

\bibitem{SalTra20}
H Salje, et~al., {Estimating the burden of SARS-CoV-2 in France}.
\newblock {\em\protect\textit{Science}} \textbf{369}, Issue 6500, 208--211 (2020).

\bibitem{GatBer20}
M Gatto, et~al., Spread and dynamics of the COVID-19 epidemic in Italy: Effects
  of emergency containment measures.
\newblock {\em\protect\textit{Proceedings of the National Academy of
  Sciences}} \textbf{117}, 10484--10491 (2020).


\bibitem{GarCha20}
AL Garc{\'\i}a-Basteiro, et~al., Monitoring the COVID-19 epidemic in the
  context of widespread local transmission.
\newblock {\em\protect\textit{The Lancet Respiratory Medicine}}
  \textbf{8}, 440--442 (2020).

\bibitem{RoqKle20b}
L Roques, EK Klein, J Papaix, A Sar, S Soubeyrand, {Impact of lockdown on the
  epidemic dynamics of COVID-19 in France}.
\newblock {\em\protect\textit{Frontiers in Medicine}} https://doi.org/10.3389/fmed.2020.00274 (5 June 2020).

\bibitem{ZhaLit20}
J Zhang, et~al., Age profile of susceptibility, mixing, and social distancing
  shape the dynamics of the novel coronavirus disease 2019 outbreak in China.
\newblock {\em\protect\textit{medRxiv}} https://dx.doi.org/10.1101\%2F2020.03.19.20039107 (20 March 2020).

\bibitem{Ber03}
LM Berliner, {Physical-statistical modeling in geophysics}.
\newblock {\em\protect\textit{{J Geophys Res}}} \textbf{{108}}, {8776}
  (2003).

\bibitem{SouLai09}
S Soubeyrand, AL Laine, I Hanski, A Penttinen, Spatiotemporal structure of
  host-pathogen interactions in a metapopulation.
\newblock {\em\protect\textit{American Naturalist}} \textbf{174},
  308--320 (2009).

\bibitem{ZhoYu20}
F Zhou, et~al., {Clinical course and risk factors for mortality of adult
  inpatients with COVID-19 in Wuhan, China: a retrospective cohort study}.
\newblock {\em\protect\textit{The Lancet}} \textbf{395}, Issue 10229, 1054--1062
 (2020).

\bibitem{HeLau20}
X He, et~al., Temporal dynamics in viral shedding and transmissibility of
  {COVID-19}.
\newblock {\em\protect\textit{Nature Medicine}}, \textbf{26},
  Issue 5, 672--675 (2020).

\bibitem{Ken57}
D Kendall, {Discussion of 'Measles periodicity and community size' by MS
  Bartlett}.
\newblock {\em\protect\textit{J. Roy. Stat. Soc. A}} \textbf{120}, 64--76
  (1957).

\bibitem{BonBer18}
L Bonnasse-Gahot,  H Berestycki, M-A Depuiset, M B Gordon, S Roch{\'e}, N Rodriguez, J-P Nadal, Epidemiological modelling of the 2005 {F}rench riots:
  a spreading wave and the role of contagion.
\newblock {\em\protect\textit{Scientific Reports}} \textbf{8}, 1--20
  (2018).

\bibitem{BroHuf06}
D Brockmann, L Hufnagel, T Geisel, The scaling laws of human travel.
\newblock {\em\protect\textit{Nature}} \textbf{439}, 462--465 (2006).

\bibitem{MeyHel14}
S Meyer, L Held, , et~al., Power-law models for infectious disease spread.
\newblock {\em\protect\textit{The Annals of Applied Statistics}}
  \textbf{8}, 1612--1639 (2014).


\bibitem{Par1896}
V Pareto, Cours d'Economie Politique, (Lausanne: F. Rouge, 1896).

\bibitem{GauGha10}
J Gaudart, et~al., Demography and diffusion in epidemics: malaria and black
  death spread.
\newblock {\em\protect\textit{Acta Biotheoretica}} \textbf{58}, 277--305
  (2010).

\bibitem{Aka74}
H Akaike, A new look at the statistical model identification.
\newblock {\em\protect\textit{IEEE transactions on automatic control}}
  \textbf{19}, 716--723 (1974).

\bibitem{Sch78}
G Schwarz, Estimating the dimension of a model.
\newblock {\em\protect\textit{Ann Stat}} \textbf{6}, 461--464 (1978).

\bibitem{NisGer09}
H Nishiura, G Chowell, The effective reproduction number as a prelude to
  statistical estimation of time-dependent epidemic trends in {\em Mathematical
  and statistical estimation approaches in epidemiology}.
\newblock (Springer), pp. 103--121 (2009).


\bibitem{DonDu20}
E Dong, H Du, L Gardner, {An interactive web-based dashboard to track COVID-19
  in real time}.
\newblock {\em\protect\textit{The Lancet Infectious Diseases}} \textbf{20}, Issue 5, 533--534 (2020).


\bibitem{SouRib20}
S Soubeyrand, et~al., {The current COVID-19 wave will likely be mitigated in
  the second-line European countries}.
\newblock {\em\protect\textit{medRxiv}} https://doi.org/10.1101/2020.04.17.20069179 (22 April 2020).

\bibitem{FonTon20}
A Fontanet, et~al., {Cluster of COVID-19 in northern France: A retrospective
  closed cohort study}.
\newblock {\em\protect\textit{medRxiv}} https://doi.org/10.1101/2020.04.18.20071134 (23 April 2020).

\bibitem{PolPer20}
M Poll{\'a}n, et~al., {Prevalence of SARS-CoV-2 in Spain (ENE-COVID): a
  nationwide, population-based seroepidemiological study}.
\newblock {\em\protect\textit{The Lancet}} https://doi.org/10.1016/S0140-6736(20)31483-5 (6 July 2020).

\bibitem{ChiDav20}
M Chinazzi, et~al., The effect of travel restrictions on the spread of the 2019
  novel coronavirus ({COVID-19}) outbreak.
\newblock {\em\protect\textit{Science}} \textbf{368}, 395--400 (2020).

\bibitem{LiuZha20}
X Liu, S Zhang, Covid-19: Face masks and human-to-human transmission.
\newblock {\em\protect\textit{Influenza and Other Respiratory Viruses}}
  (2020).


\end{thebibliography}

\clearpage

\section*{Supplementary Material}
 \renewcommand*{\thefigure}{S\arabic{figure}}
\setcounter{figure}{0}

\section*{SI 1. Supplementary figures}

\begin{figure}[h!]
\center
%\graphicspath{{images/}}
\includegraphics[width=0.48\textwidth]{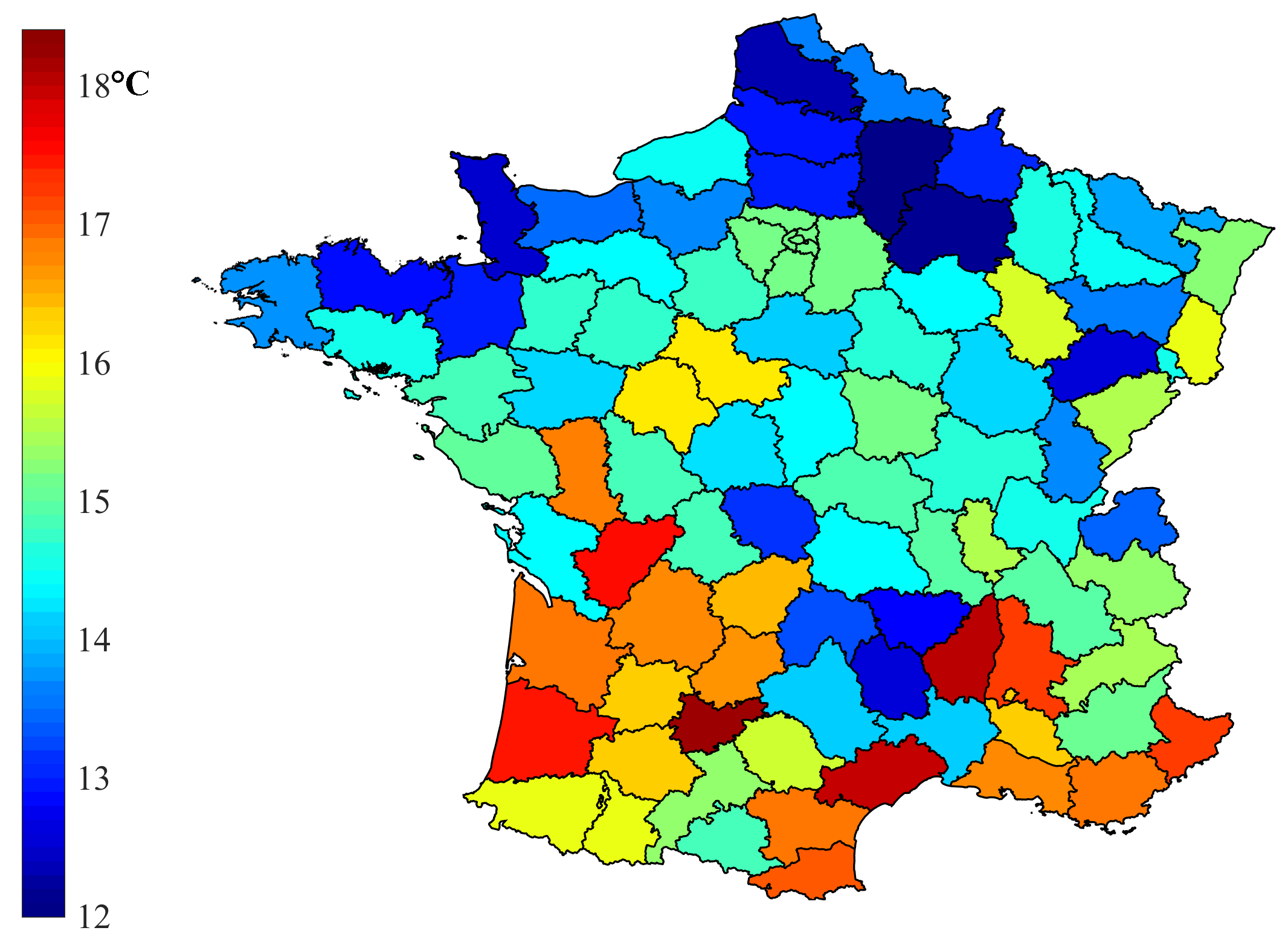}
\includegraphics[width=0.48\textwidth]{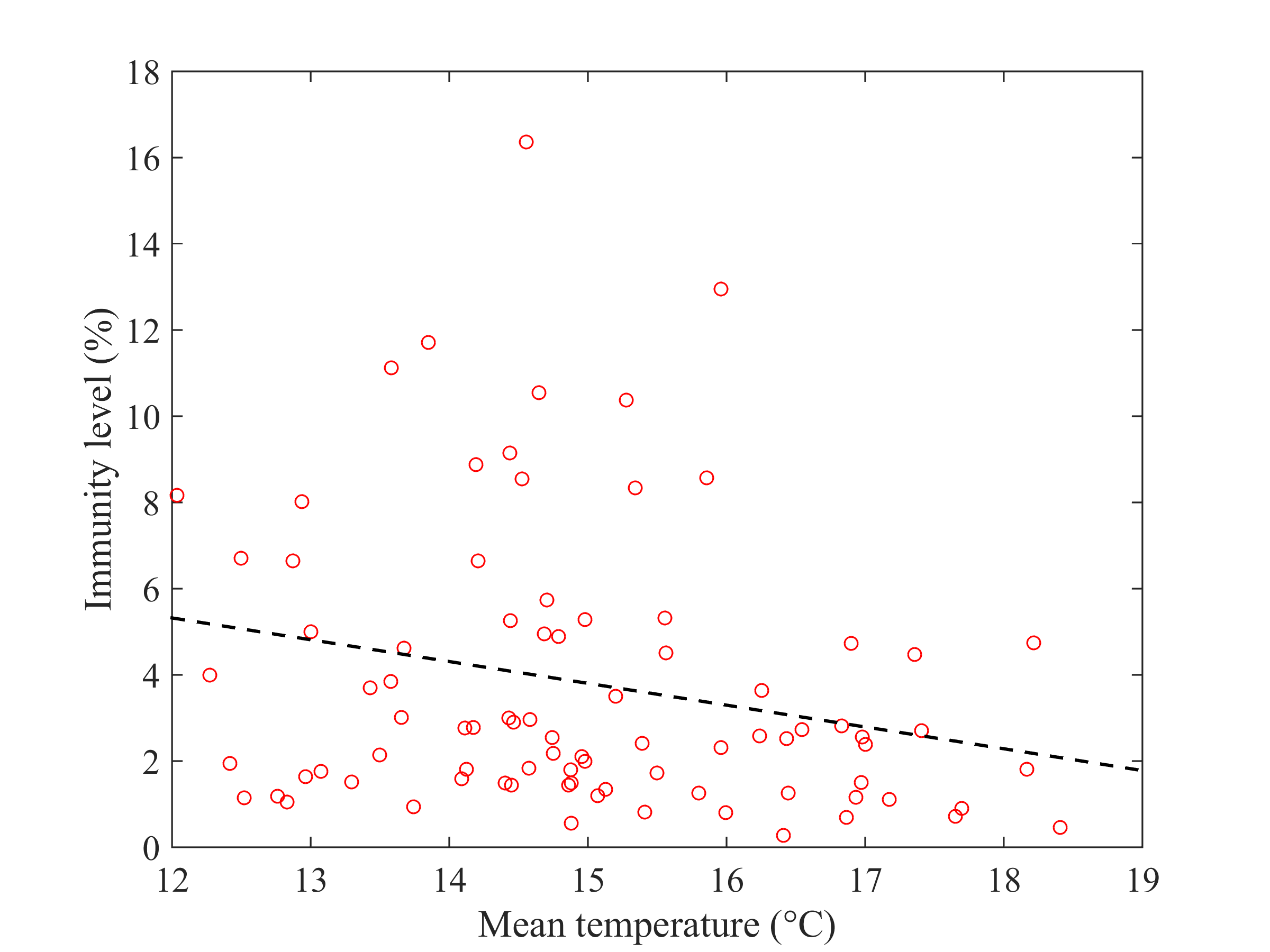}
\caption{(a) Mean temperature in each county over the period ranging from 30 March 2020 to 11 June 2020. (b) Immunity rate vs mean temperature: we observe a  negative correlation between the mean temperature and the immunity rate (Pearson correlation coefficient: $-0.24$).}
\label{fig:temp}
\end{figure}

\begin{figure}[h!]
\center
\includegraphics[width=0.32\textwidth]{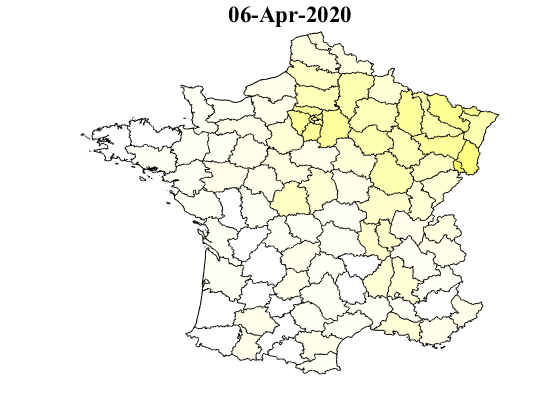}
\includegraphics[width=0.32\textwidth]{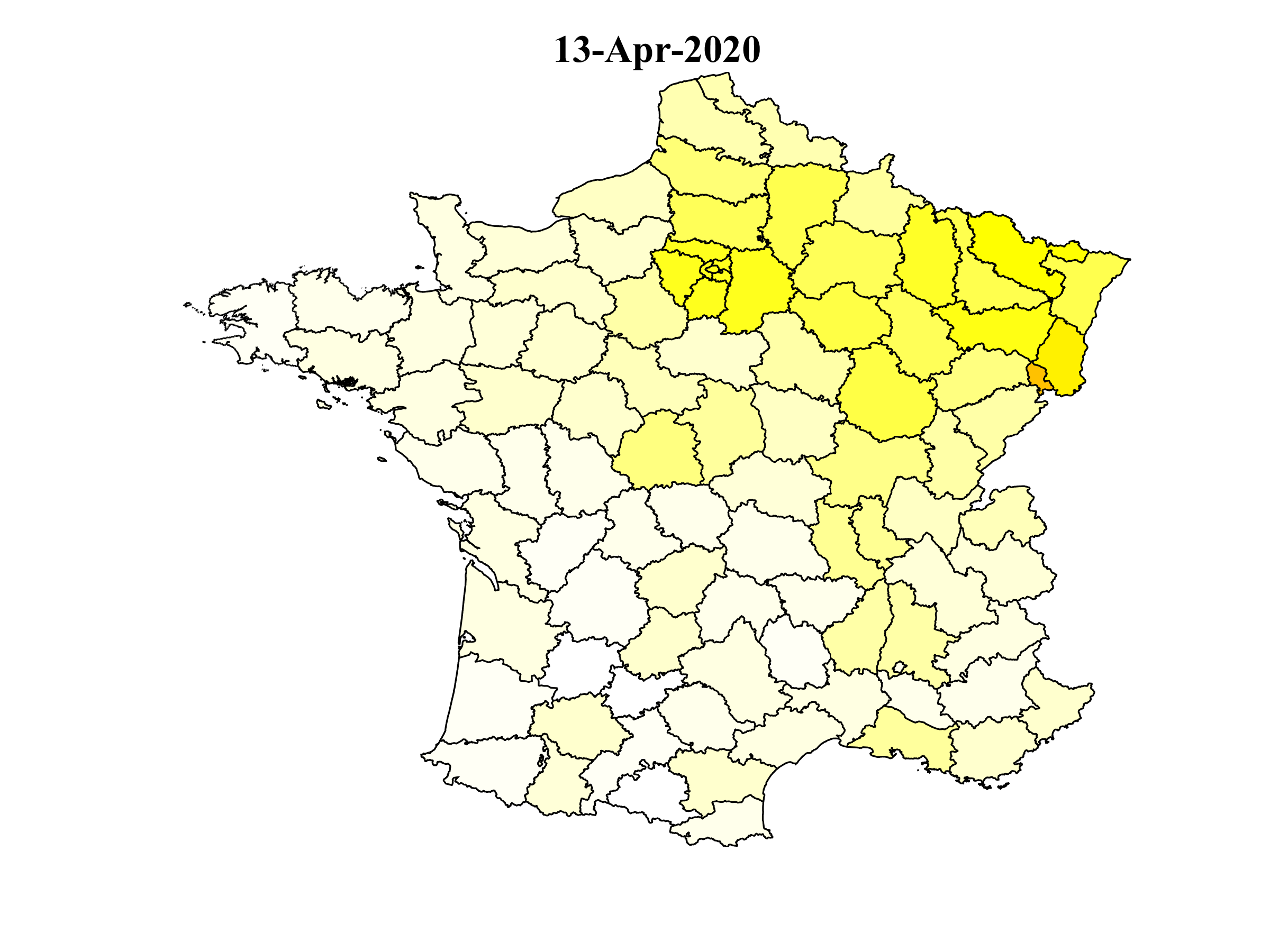}
\includegraphics[width=0.32\textwidth]{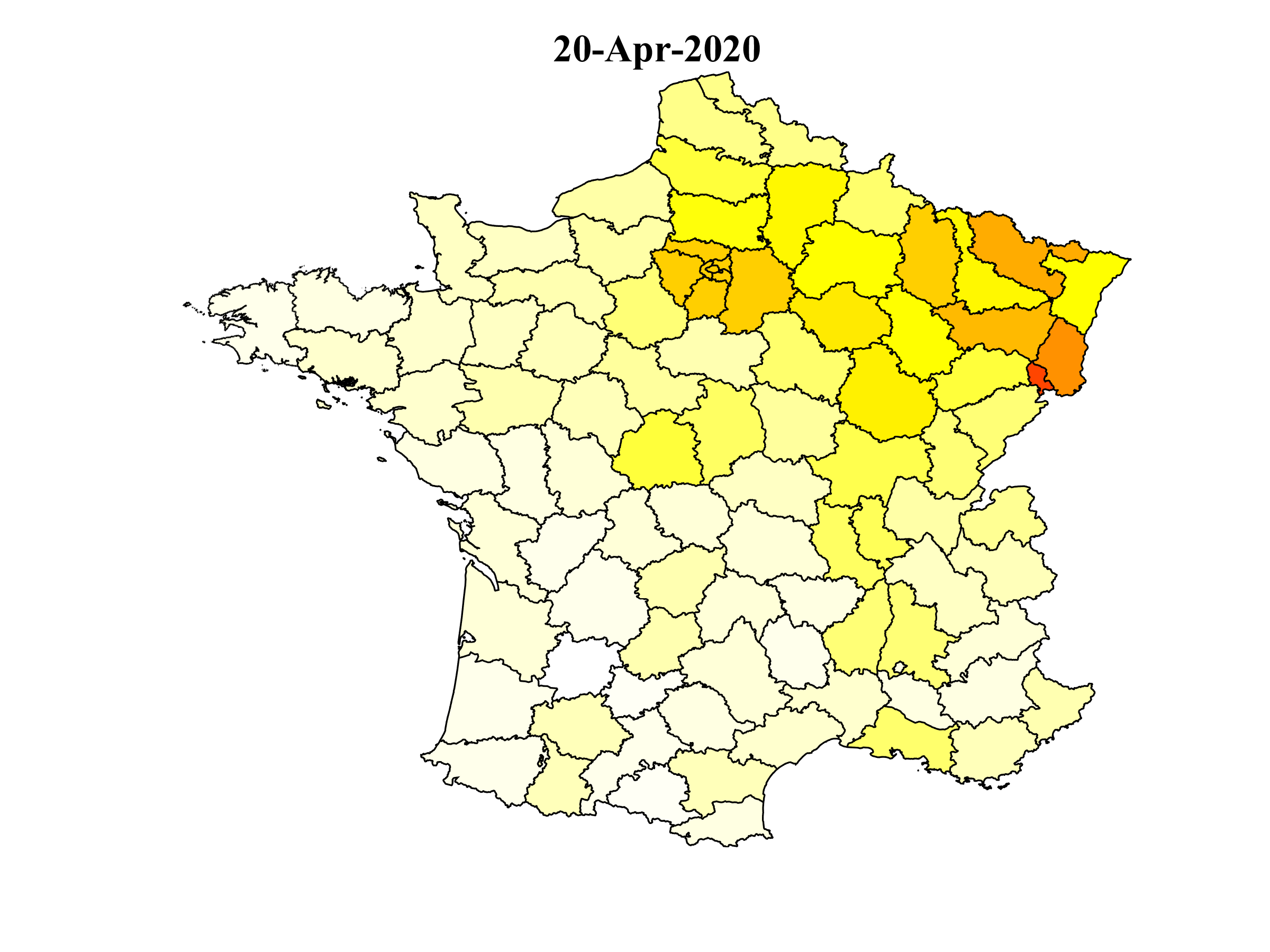}
\includegraphics[width=0.32\textwidth]{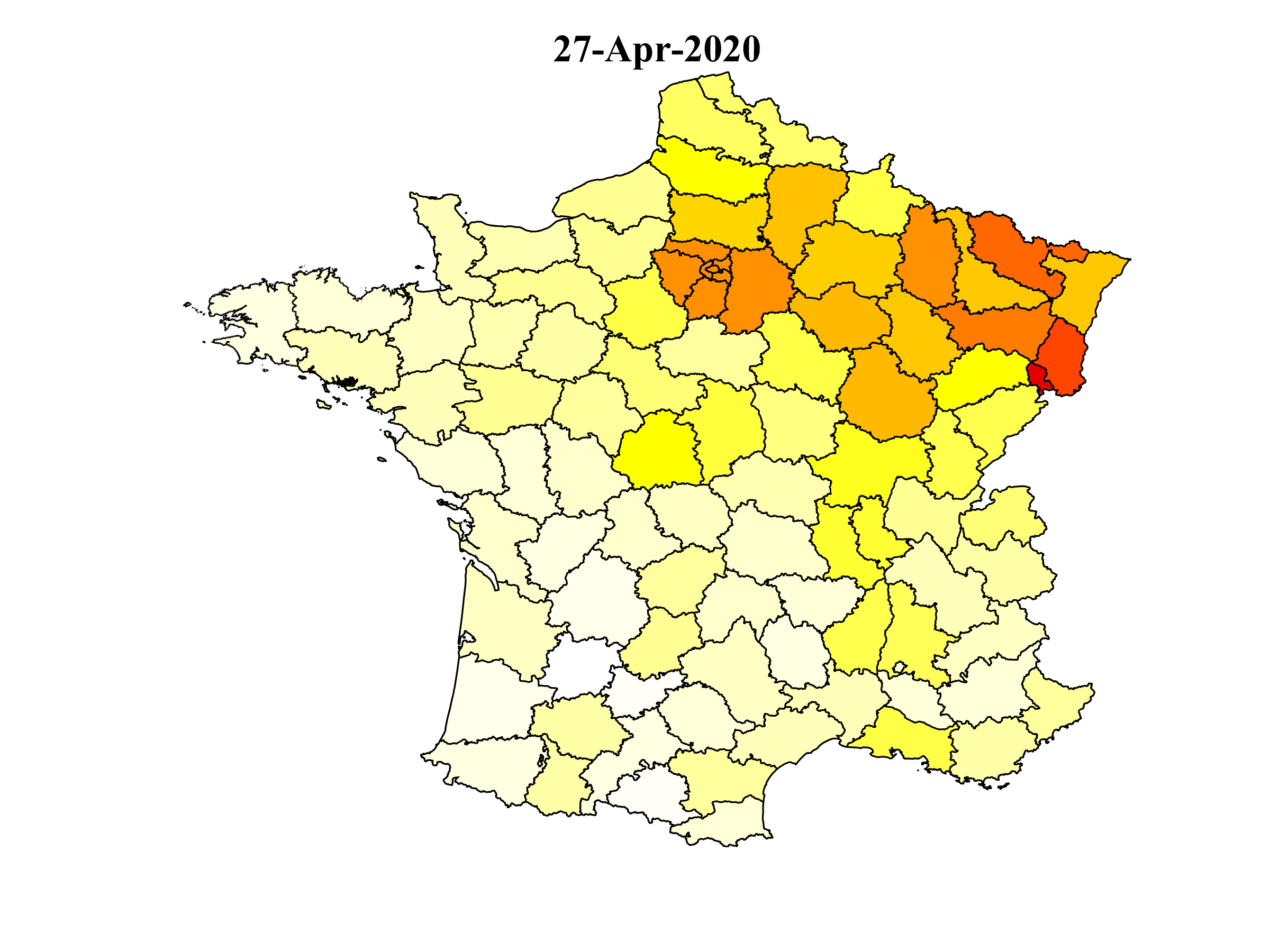}
\includegraphics[width=0.32\textwidth]{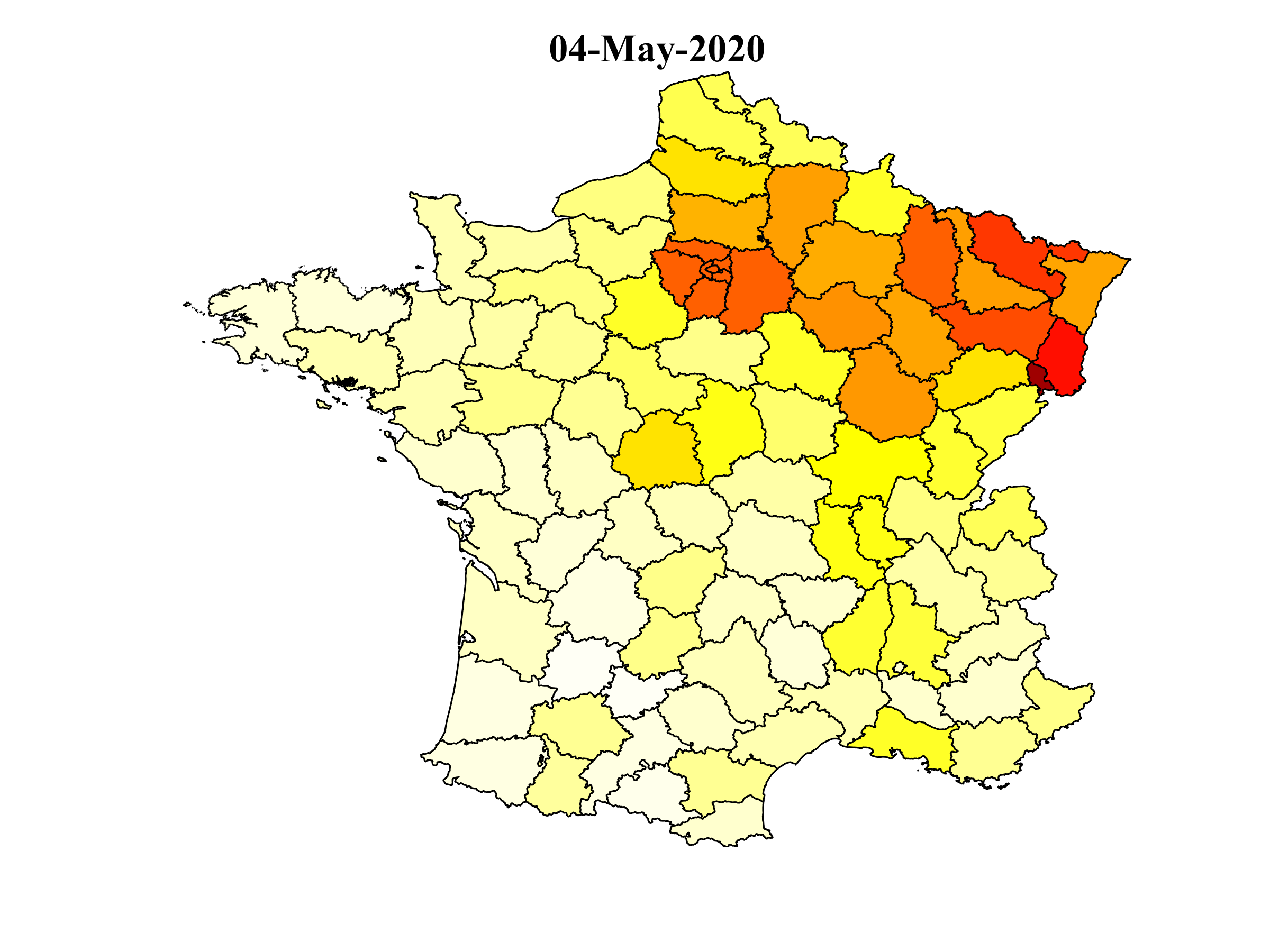}
\includegraphics[width=0.32\textwidth]{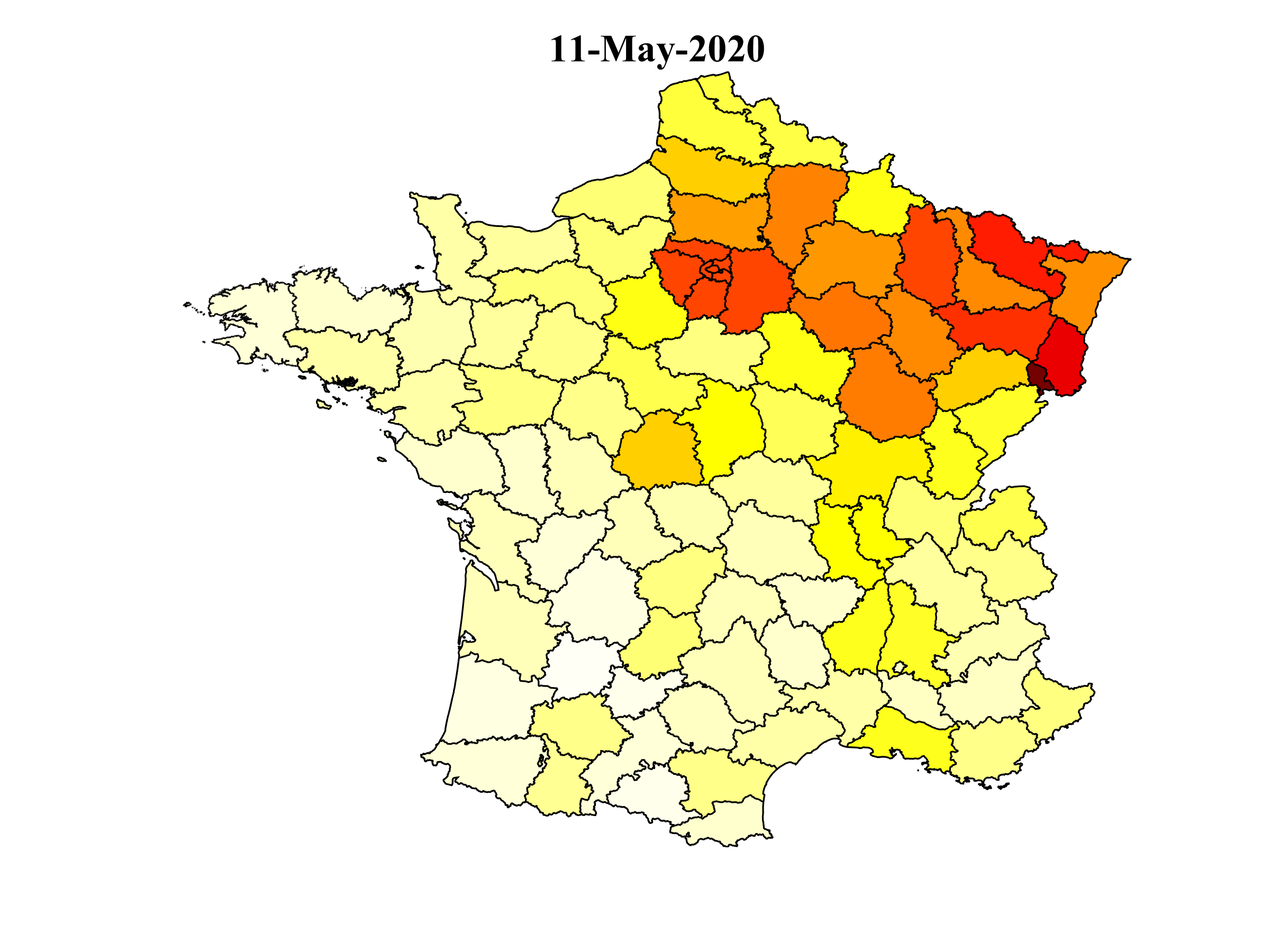}
\includegraphics[width=0.32\textwidth]{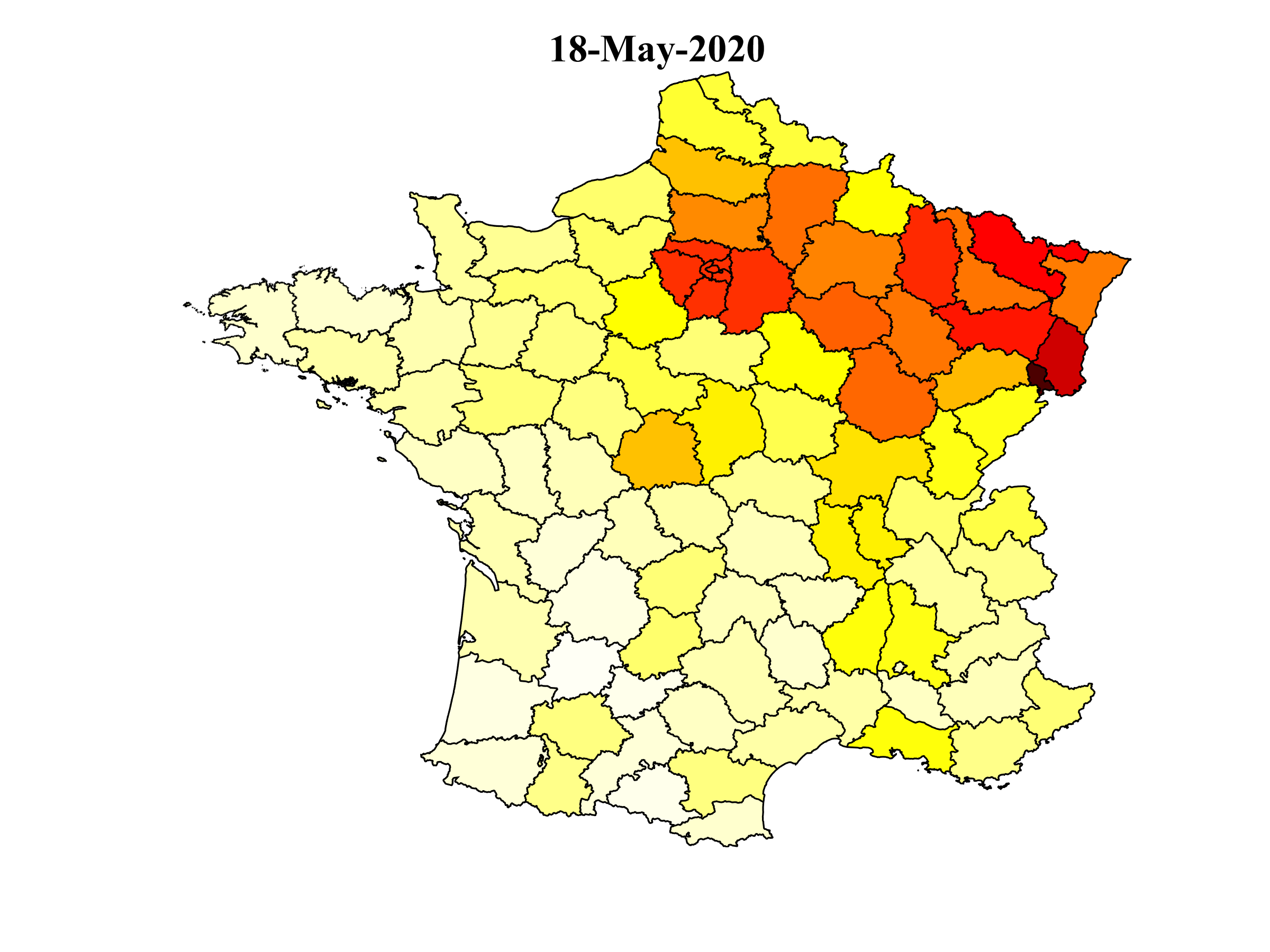}
\includegraphics[width=0.32\textwidth]{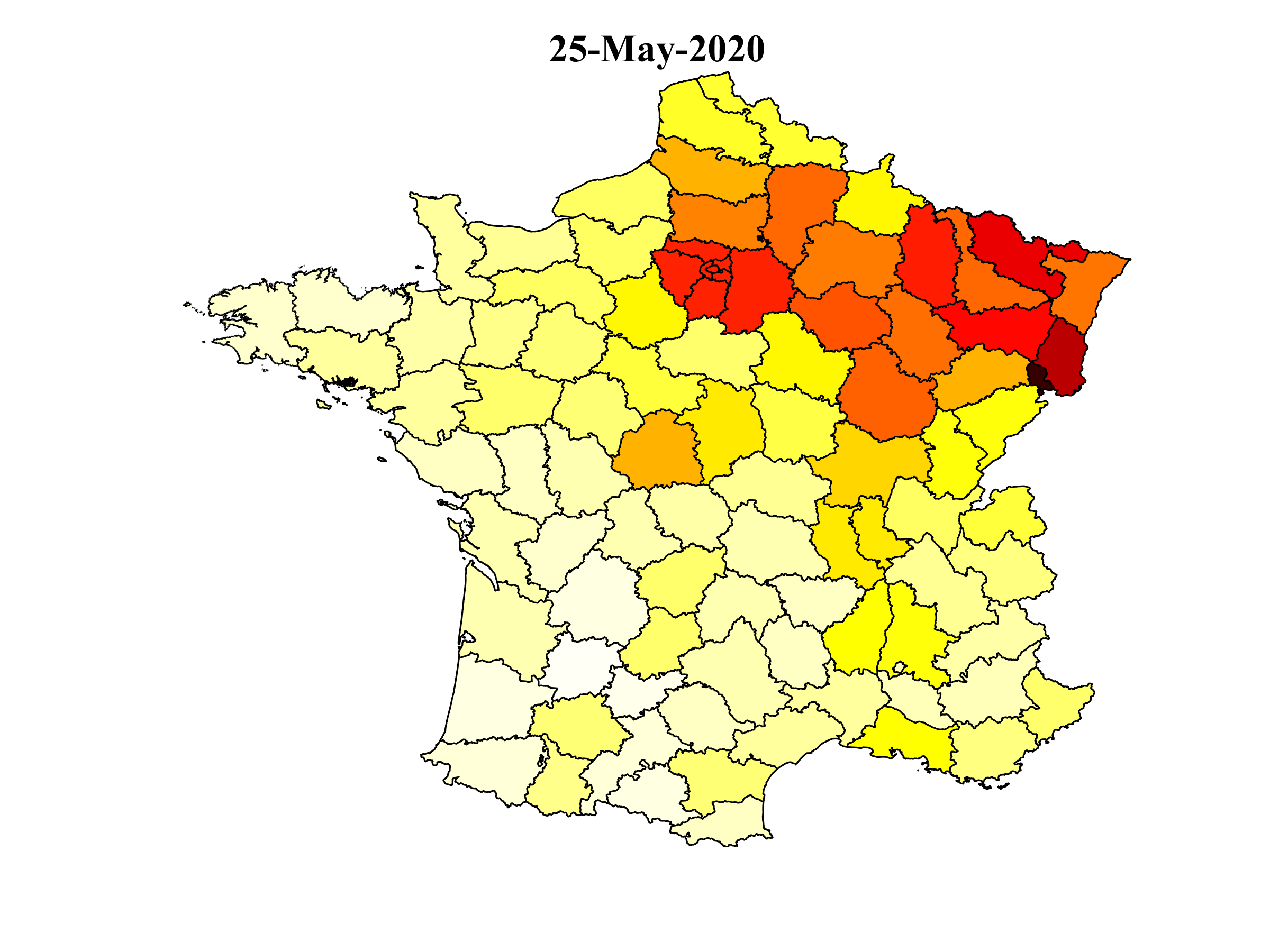}
\includegraphics[width=0.32\textwidth]{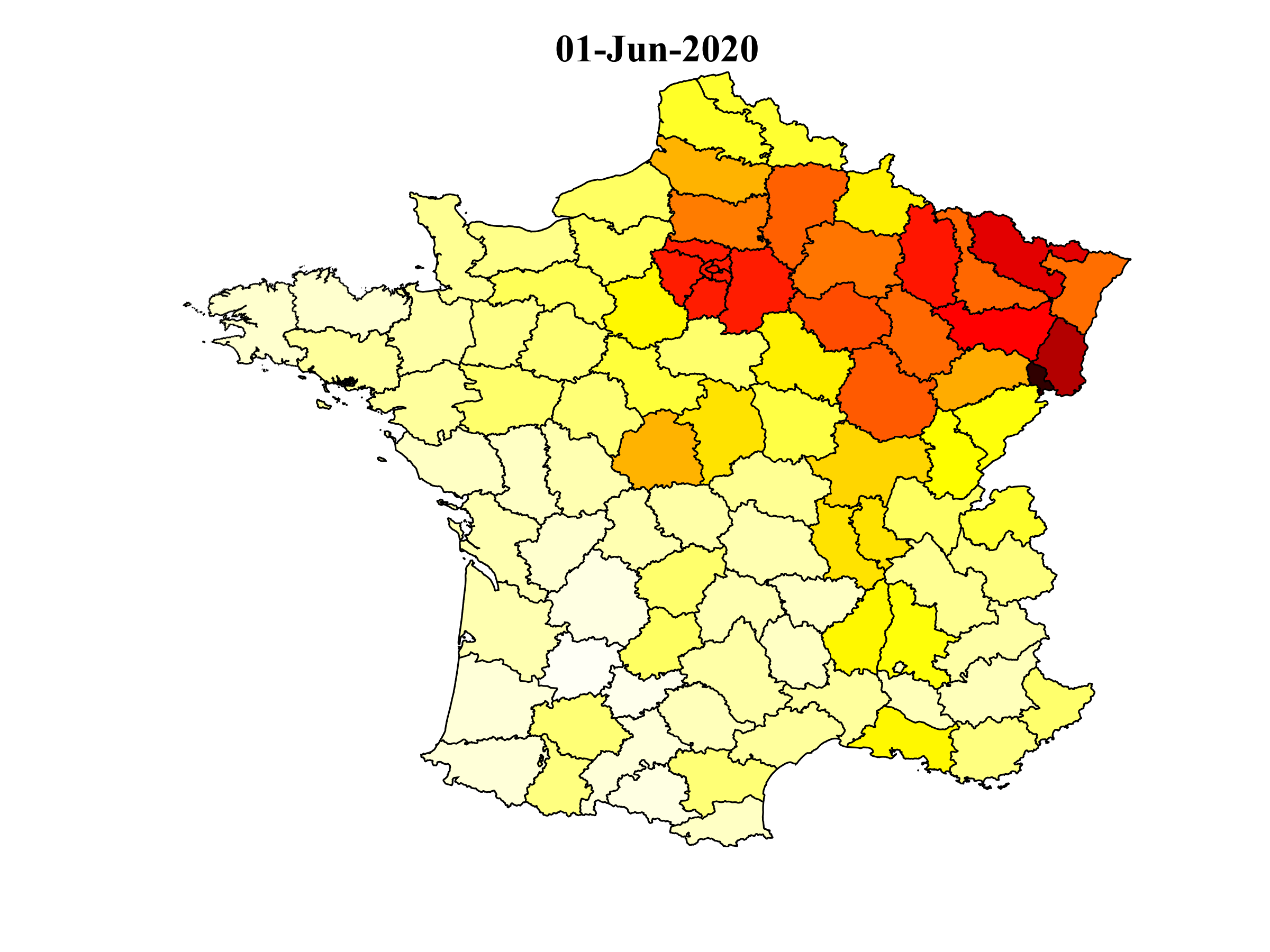}
\includegraphics[width=0.32\textwidth]{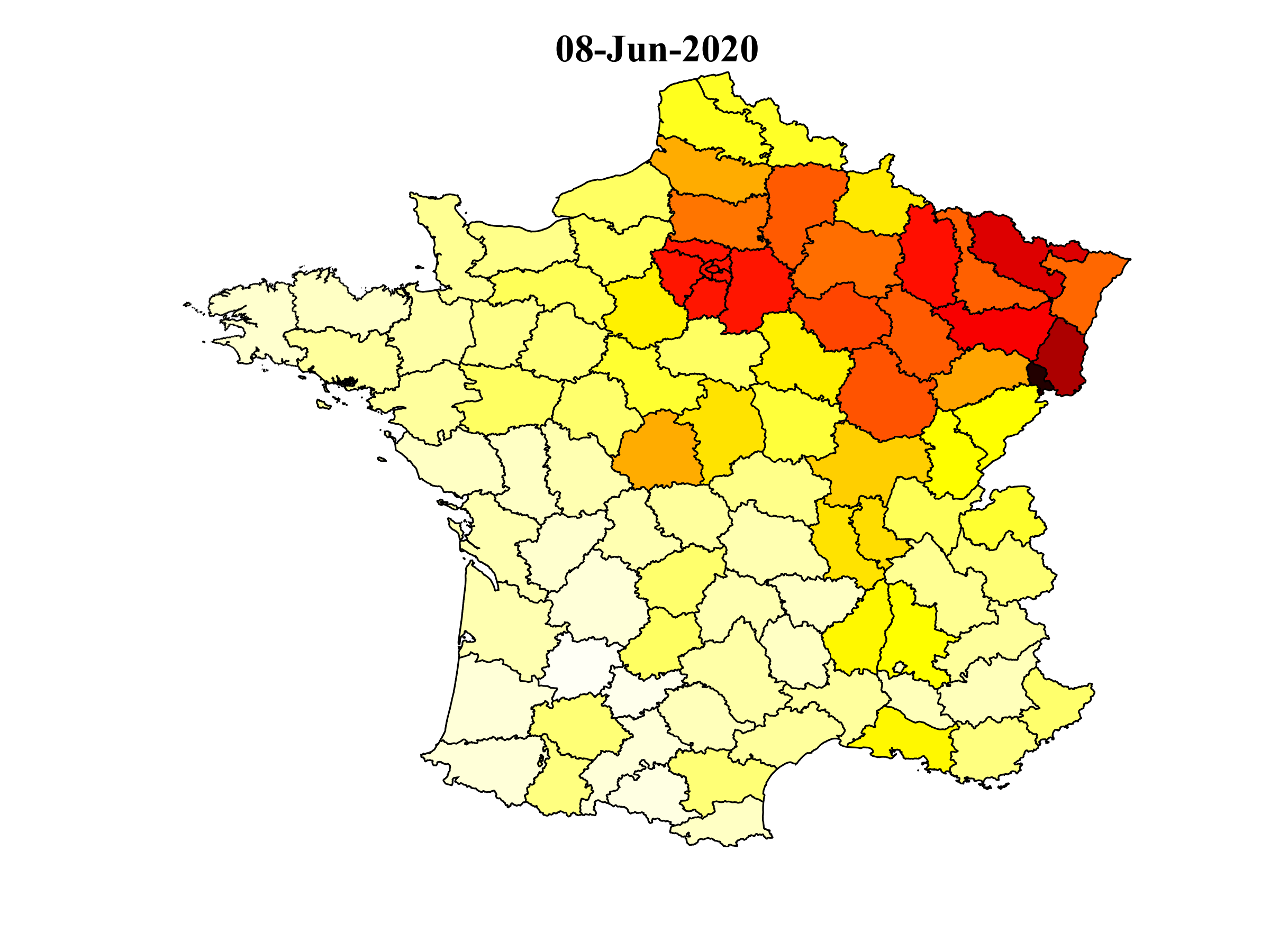}
\includegraphics[width=0.05\textwidth]{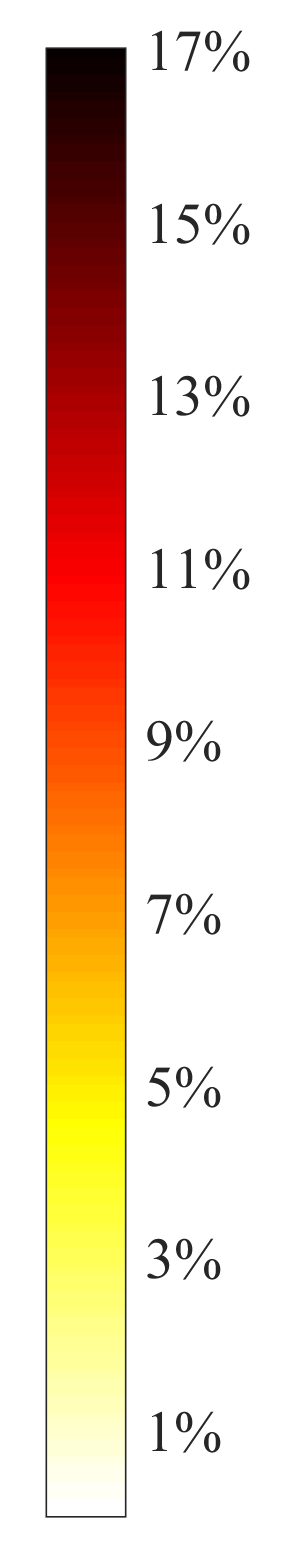}
\caption{Dynamics of the immunity rate given by model $\M_3$, from 6 April 6 to 8 June.}
\label{fig:timeline}
\end{figure}

\begin{itemize}
    \item The mean temperatures during the observation period are presented in Fig.~\ref{fig:temp}, together with a scatter plot of the mean temperature vs immunity rate.

    (source: https://www.data.gouv.fr/fr/datasets/temperature-quotidienne-departementale-depuis-janvier-2018/).
    \item In Fig.~\ref{fig:timeline}, we describe the timeline of the spatio-temporal dynamics of the immunity rate during the observation period.
\end{itemize}

\clearpage

\section*{SI 2. Auxiliary models}
At  each date $t$, the time-dependent parameters  $\alpha(t)$ in $\M_{0}$ and $\M_{1}$ and $\alpha_k(t)$ in $\M_{2}$ maximum likelihood estimators (MLEs) in the window $(t-\tau/2,t+\tau/2)$ of the following auxiliary models:
\begin{equation}\label{eq:Modab} \tag{$\tilde{\M}_{0,t}$}
\baco{l}
 \ds \tS'(s)=- \frac{\tA}{N} \, \tS \, \tI, \vspace{1mm}\\
 \ds \tI'(s)= \frac{\tA}{N} \, \tS \, \tI - (\beta+\gamma) \, \tI, \vspace{1mm}\\
 \ds \tR'(s)=\beta \,  \tI,\vspace{1mm}\\
 \ds \tD'(s)=\gamma \, \tI,
\eaco \hbox{ for }s\in (t-\tau/2,t+\tau/2),
\end{equation}
and
\begin{equation}\label{eq:Mod0b} \tag{$\tilde{\M}_{1,t}$}
\baco{l}
 \ds \tS_k'(s)=- \frac{\tA}{N_k} \, \tS_k \, \tI_k, \vspace{1mm}\\
 \ds \tI_k'(s)= \frac{\tA}{N_k} \, \tS_k \, \tI_k - (\beta+\gamma) \, \tI_k, \vspace{1mm}\\
 \ds \tR_k'(s)=\beta \,  \tI_k,\vspace{1mm}\\
 \ds \tD_k'(s)=\gamma \, \tI_k,
\eaco \hbox{ for }s\in (t-\tau/2,t+\tau/2),
\end{equation}
and
\begin{equation}\label{eq:Mod1b} \tag{$\tilde{\M}_{2,t}$}
\baco{l}
 \ds \tS_k'(s)=- \frac{\tA_k}{N_k} \, \tS_k \, \tI_k, \vspace{1mm}\\
 \ds \tI_k'(s)= \frac{\tA_k}{N_k} \, \tS_k \, \tI_k - (\beta+\gamma) \, \tI_k, \vspace{1mm}\\
 \ds \tR_k'(s)=\beta \,  \tI_k,\vspace{1mm}\\
 \ds \tD_k'(s)=\gamma \, \tI_k,
\eaco \hbox{ for }s\in (t-\tau/2,t+\tau/2).
\end{equation}
The initial condition in these models is computed iteratively from the solutions of $\M_{0}$, $\M_{1}$ and $\M_{2}$, respectively, over the period $[t_i,t-\tau/2]$.

\end{document}